\documentclass[11pt,onecolumn]{IEEEtran}

\usepackage{epsfig,amsmath,amssymb,amsbsy,amsthm,epsf,amsthm,scalefnt,subfig,multirow,algorithm,algorithmic,mathtools}
\usepackage{xcolor}
\usepackage{float}
\usepackage{cite}

\def\inf{\mathop{\mathrm{inf}}}

\def\b0{{\pmb{0}}} 

\newcommand{\hsppp}{{\hspace{0.02in}}}
\newcommand{\hspp} { {\hspace{0.05in}} }
\newcommand\ignore[1]{}

\newcommand\myeq{\mathrel{\overset{\makebox[0pt]{\mbox{\normalfont\tiny\sffamily i.i.d.}}}{\sim}}}

\begin{document}

\title{Multi-Sensor Sequential Change Detection with Unknown Change Propagation Pattern}

\author{Mehmet Necip~Kurt
        and~Xiaodong~Wang
\thanks{The authors are with the Department
of Electrical Engineering, Columbia University, New York, NY 10027, USA (e-mail: m.n.kurt@columbia.edu; wangx@ee.columbia.edu).}
}

\maketitle

\begin{abstract}
The problem of detecting changes with multiple sensors has received
significant attention in the literature. In many practical applications
such as critical infrastructure monitoring and modeling
of disease spread, a useful change propagation model
is one where change eventually happens at all sensors, but where
not all sensors witness change at the same time instant. While
prior work considered the case of known change
propagation dynamics, this paper studies a more general setting of unknown change
propagation pattern (trajectory). A Bayesian formulation of the problem
in both centralized and decentralized settings is studied with the goal
of detecting the first time instant at which any sensor witnesses a change.
Using the dynamic programming (DP) framework, the optimal solution structure
is derived and in the rare change regime, several more practical change detection algorithms are proposed. Under certain conditions,
the first-order asymptotic optimality of a proposed algorithm called multichart test is shown
as the false alarm probability vanishes. To further reduce the computational complexity, change detection algorithms are proposed based on online estimation of the unknown change propagation pattern. Numerical studies illustrate
that the proposed detection techniques offer near-optimal performance. Further, in the
decentralized setting, it is shown that if an event-triggered sampling scheme
called level-crossing sampling with hysteresis (LCSH) is used for sampling and transmission of local statistics,
the detection performance can be significantly improved using the same amount of communication resources compared to the conventional uniform-in-time sampling (US) scheme.
\end{abstract}


\section{Introduction} \label{sec:intro}

Over the last decade, significant engineering and economic challenges have been overcome in the development of sensor technology. In particular, with low cost, low energy consumption and higher reliability, a large number of sensors can now be used for monitoring critical infrastructures such as bridges, water pipes, roadways, oil and gas networks, etc. Other scenarios where multiple sensors are deployed include monitoring remote/dangerous environments in industrial settings, environmental applications, and battlefield applications, etc. {Networked sensor systems are commonly used for situation assessment, more particularly detection/estimation of an event/quantity of interest, over surveillance areas \cite{Niu08,Ciuonzo17,Ciuonzo11,Chair86,Vempaty14,YilmazTAES}.}

{A typical task of sensor networks is to monitor the underlying stochastic process associated with the application and to infer any deviation from normal behavior which can be attributed to unforeseen changes in the system/environment. Particular applications are real-time detection of anomalies such as solar flares (big energy releases from the Sun) that affects the long-range radio communications \cite{chen2016,Xie13}, cyber-attacks targeting the power grids \cite{Necip18,Necip18b,Li_15,Necip19,Necip19b}, and faulty sensors in aircraft systems, inertial navigation systems (of e.g., planes, boats, and rackets), and the critical infrastructures \cite{CriticalStructures,Noel17,chen2016,Raghavan10,Basseville93}. For instance, in the inertial navigation systems, rapid detection of faulty sensors is crucial to avoid using faulty measurements in the navigation equations \cite{Basseville93}.}

In sequential change detection problems, an abrupt change occurs in the sensing environment at some unknown time and statistical properties of the observed process alter. The goal is to detect the change as soon as possible after it occurs while limiting the risk of false alarm. Statistical inference about the underlying process is typically done through measurements taken sequentially over time. At each time, a decision maker either declares a change or continues to have further measurements in the next time interval. The design goal is to optimally balance the detection delay and the false alarm to have a timely and accurate response to the changes.

In the literature, the change-point is either considered as a deterministic unknown quantity or a random variable with a known geometric distribution. The first approach corresponds to a minimax formulation in which a worst-case detection delay is minimized subject to a lower bound on the mean time between false alarms \cite{Lorden71,Pollak85}. The other approach corresponds to a Bayesian formulation in which the expected detection delay is minimized subject to an upper bound on the false alarm probability \cite{Shiryaev78}. For a detailed review of the sequential change detection formulations, algorithms, performance analyses, and applications, please see \cite{Poor08,Basseville93,Veeravalli14,Polunchenko12}.

\subsection{Literature Survey on Multi-Sensor Change Detection}

Detecting changes with multiple sensors has been a problem of growing interest and a number of works have addressed this problem. If a decision maker has all the system-wide information, then the problem is called a centralized change detection problem. If sensors have some communication constraints, then measurements collected by sensors cannot be directly sent to a central processor (fusion center) and the problem is called a decentralized change detection problem, which is introduced in \cite{Veeravalli01} under a Bayesian setting where the objective is the joint optimization of sensor messages and the fusion center policy.

Two approaches are commonly employed in decentralized change detection with a fusion center: (i) each sensor makes local change detection based on its own measurements and sends its decision to the fusion center and (ii) each sensor computes a local statistic based on its measurements and sends a quantized version of the local statistic to the fusion center. In \cite{Tartakovsky08}, both scenarios are studied in both Bayesian and minimax settings assuming the same change-point for all sensors and asymptotically optimal change detection schemes are proposed. Moreover, the detection performance is shown to be better in (ii) compared to (i) in general.

Most of the existing studies either consider the scenario where change happens at the same time-instant across all sensors or when only a random subset of sensors witness change, also at the same time-instant. In \cite{Tartakovsky05}, a generalization of the single sensor-optimal Shiryaev procedure is proposed, and it is shown that the detection delay of the multi-sensor procedure scales linearly with the number of sensors making observations. In hindsight, this conclusion is natural and not entirely surprising since the sensors witness change at the same time-instant and offer observational diversity in the decision-making process. The minimax version of this problem is studied in \cite{Mei05} where near-optimal procedures from the single sensor setting such as the Cumulative Sum (CUSUM) and Shiryaev-Roberts procedures are extended to the multi-sensor setting. The performance analysis of these procedures illustrate the same trend(s) as the Shiryaev procedure in the Bayesian setting.

The case in which an unknown subset of sensors simultaneously observe the change and the other sensors are not affected by the change corresponds to  so-called multichannel change detection problems \cite{Tartakovsky02,Mei10,Mei11,Siegmund13,Banerjee15,Fellouris16}. In \cite{Mei10}, an asymptotically optimal scheme is proposed in which each sensor sends its local CUSUM statistics to the fusion center which computes the sum of the local statistics as the decision statistic. It is discussed that using global information obtained through all sensors leads to lower detection delays compared to using only local information and enables to control the false alarm rate of the overall system. Further, since the information obtained from sensors that do not observe the change produces a noise effect on the decision-making process, in \cite{Mei11}, only local statistics above some thresholds are considered in decision-making process to improve the detection performance. With the same purpose of suppressing this noise effect, a prior probability representing the proportion of sensors observing the change among all sensors is incorporated into the formulation in \cite{Siegmund13}. Furthermore, a decentralized change detection problem is considered in \cite{Banerjee15} where costs of taking measurements and the communication between sensors and the fusion center are incorporated into the formulation and some resource-effective asymptotically optimal change detection algorithms are proposed in a minimax setting.

In a similar set of studies, it assumed that change occurs due to an unknown source of change at an unknown time. The goal is to jointly detect the change time and identify cause of the change. Hence, in addition to false alarm, false identification is also incorporated into the formulation and controlled by the fusion center. Such problems are called change detection and isolation problems that are introduced in \cite{Nikiforov95} in a minimax setting. In \cite{Geng13}, a Bayesian version of the problem is studied and a distance-based sensing model is assumed where each sensor can observe the change only if the change occurs in its limited sensing range. In \cite{Premkumar09}, a distributed version of this problem is considered in a minimax setting where each sensor is allowed to communicate only with its neighbors. In both \cite{Geng13} and \cite{Premkumar09}, the goal is to jointly detect the change time and the change location.

In many practical applications, change does not happen at the same time-instant across all sensors. For example, in an application where sensors are monitoring cracks in a bridge, the slow progression and development of the crack due to geological and chemical changes in the bridge infrastructure ensures that change is seen at different sensors at different points in time. Similarly, different members of a population encounter different exposure levels to a certain disease depending on the precise evolution of the disease-causing agent(s) over the environment, thus resulting in different onset times for a disease in any member of the population. The focus of this paper is on such applications, where change propagates across the sensors, one sensor at a time.

In \cite{Raghavan10}, a Bayesian centralized change detection problem is considered where change propagates across the deployed sensor set via a known change process. A first-order asymptotically optimal detection procedure that leverages the known change dynamics is proposed for detecting the change-point and the proposed procedure is shown to significantly improve performance over a mismatched procedure that does not leverage the change dynamics. In \cite{Lifeng13}, \cite{Raghavan10} is extended to a setting where the sensor first observing the change is assumed to be unknown. False identification probability of this sensor is also incorporated into the formulation, however, given the sensor observing the change first, a predetermined order of observing the change among sensors is assumed. In \cite{Ludkovski12}, a wave propagation model is considered for the change signal where the signal, having an unknown origin and radial velocity, gradually spreads through space and each sensor observes the change when the signal reaches the sensor location. A continuous-time Bayesian centralized formulation is considered and numerical approximation methods to implement the optimal solution are provided. While the change propagation models assumed in \cite{Raghavan10,Lifeng13,Ludkovski12} are of interest in many applications where the change propagation pattern can be inferred from the physical process causing the change, in general, methods that do not assume the knowledge of the change pattern are important.

Having different change points for different sensors is also considered in \cite{Hadjiliadis09,Zhang14} without assuming any change propagation mechanism in a minimax setting. Asymptotically optimal detection schemes are proposed where each sensor makes a local change detection and communicates with the fusion center only once (one-shot scheme) after it detects the change and the fusion center declares the change after the first received message from any sensor. However, due to observational diversity, using global information obtained through all sensors may lower the detection delays compared to using only local information obtained at a single sensor. 

\subsection{Contributions}

In this paper, we consider a Bayesian discrete-time multi-sensor sequential change detection problem assuming an unknown change propagation pattern among sensor nodes in both centralized and decentralized settings. We firstly obtain the optimal solution structures using the dynamic programming (DP) arguments as in \cite{Veeravalli01} and \cite{Raghavan10}. Since (i) further analysis on the optimal solutions is difficult, (ii) implementing the optimal solutions requires a computationally challenging numerical method, and (iii) the decentralized optimal solution corresponds to a practically infeasible scheme due to communication constraints in a typical sensor network, more practical {novel} algorithms are proposed for a special case where the change event happens rarely. Since the proposed algorithms are still computationally intensive to implement in real-time settings especially for large networks consisting of many sensors, a {novel} computationally efficient joint detection and estimation scheme is proposed where the change detection is performed based on the online estimation of the change propagation pattern.

Two approaches are taken to tackle the unknown change propagation pattern: it is either considered as a random variable with a known prior (uniform) distribution or a {totally unknown} deterministic quantity. For the latter case, under certain conditions on Kullback-Leibler (K-L) information distance between the post- and pre-change distributions, a first-order asymptotic optimality result is provided as the false alarm probability vanishes. The performance improvement obtained with the proposed tests over some existing tests in the literature is shown via simulations. Moreover, with the purpose of further improving the performance of the proposed decentralized detection schemes, an event-triggered sampling scheme is proposed for sampling and transmission of local statistics and its advantages over the conventional uniform-in-time sampling (US) scheme are illustrated via simulations.

\subsection{Organization}

The system model and the problem formulation are given in Sec.~\ref{sec:model}. The DP formulation and some useful results for the finite-horizon case are provided in Sec.~\ref{sec:dp}. The optimal solution structure is presented in Sec.~\ref{sec:inf_horizon}. Sec.~\ref{sec:rare_change} discusses the rare change regime where some more practical change detection algorithms are proposed and analyzed. The decentralized implementation using LCSH scheme is discussed in Sec.~\ref{sec:event_based}. The computationally efficient change detection scheme based on online estimates of the change pattern is explained in Sec.~\ref{sec:online_estimate}. Numerical results are presented in Sec.~\ref{sec:sim} to illustrate the advantages of the proposed detection schemes. Finally, Sec.~\ref{sec:conc} concludes the paper.

We firstly study a completely Bayesian setup where the change propagation pattern is assumed to have a known prior distribution. In this setup, we obtain the optimal solution structure and also propose a more practical uniform-prior test in the special rare-change regime. We then consider a setup where the change propagation pattern is totally unknown and propose a multichart test. Then, to have a computationally more feasible solution especially for large networks, we present an online estimation mechanism for the unknown change pattern and propose an estimation-based change detection algorithm. Throughout the manuscript, we present results for the centralized and the decentralized settings simultaneously. This is because the centralized setting is a special case of the decentralized setting where the exact sensor measurements become available to the fusion center instead of quantized sensor messages. We firstly study the decentralized setting with the conventional US scheme. Then, with the goal of improving the decentralized detection performance, we discuss an event-based detection scheme.

\section{System Model and Problem Formulation}
\label{sec:model}

We consider a sequential change detection problem
where there exist random propagation delays among sensor nodes in observing a change.
To be specific, let there exist $L$ sensor nodes in the network, indexed as ${1,
2, \cdots, L}$, and $z_{k,\ell}$ denote the measurement obtained at sensor $\ell$
at discrete-time instant $k \in \mathbb{N}$. The statistic of measurements at
sensor $\ell$ undergoes a change at a random time instant, denoted as $\Gamma_{\ell}$.
We assume that the measurements are independent across the sensors. Further, at any
given sensor, they are independent and identically distributed (i.i.d.) with density
function $f_0$ and $f_1$ before and after the change, respectively. In
other words,
\begin{equation}
\label{eq:obs}
z_{k,\ell} \myeq
\left\{
\begin{array}{cc}
f_0, & {\sf{if}} ~ k < \Gamma_{\ell}, \\
f_1, & {\sf{if}} ~ k \geq \Gamma_{\ell}.
\end{array} \right.
\end{equation}
We assume that
$f_0$ and $f_1$ are common across all sensors and known
a priori. Such an assumption is reasonable since the sensors typically monitor
change engendered by an underlying physical process that occurs at a random time-instant.

We are interested in the early detection of the change event. In the special case where $\Gamma_{1} = \cdots = \Gamma_L$, i.e., the
change is observed at the same time-instant at all sensors, then this problem reduces to
the classical multi-sensor change detection problem studied in~\cite{Veeravalli01}. In this paper,
we assume that the change needs not to be observed at the same time-instant at all sensors
and it can propagate with time. Further, we make the assumption that the evolution
of the change process across the sensors (including the first sensor observing change)
is unknown. In particular, let $\Pi \triangleq \left[ \pi_1, \hsppp \cdots, \hsppp
\pi_L \right]$ denote a permutation over the index set $\{1 , \cdots, L\}$ indicating
the change propagation across sensors. In other words, ${\pi_1}$ is the first sensor
observing the change, ${\pi_2}$ is the second sensor observing the change, and so on. Thus,
we have $\Gamma_{\pi_1} \leq \cdots \leq \Gamma_{\pi_L}$. In the particular case where
$\Pi = \left[ 1, \hsppp , \cdots, \hsppp L \right]$, the problem studied here reduces to
the case where the change propagation pattern is known~\cite{Raghavan10}.

Since there are $L$ sensors in the network, $\Pi$ can take $L!$ possible
values which can be associated with a probability mass function (pmf). As an example,
we assume that the change propagation patterns are equally likely, i.e.,
\begin{gather} \label{eq:uni_perm}
\mathbb{P}(\Pi = \Pi_j) = \frac{1}{L!}, ~ j = 1, 2, \dots, L!,
\end{gather}
and we call the corresponding scheme the uniform-prior scheme. Later in Sec.~\ref{sec:rare_change} and Sec.~\ref{sec:online_estimate},
we remove this assumption and consider the change propagation pattern as an unknown non-random quantity.

\textit{Remark 1:} Since any change pattern among the $L!$ possibilities can be observed in general,
all possibilities need to be considered to determine decision statistics, that significantly increases
the computationally complexity as $L$ increases. In some special applications,
if it is known a priori that some patterns are not possible to observe, then such patterns can simply be discarded
to reduce the computational complexity. {Further in Sec.~\ref{sec:online_estimate},
we provide an online estimation mechanism for the unknown change pattern,
that considerably reduces the computational complexity of the corresponding detection scheme}.

\subsection{Change  Propagation Model}

We consider a Bayesian formulation where $\Gamma_{\pi_{\ell}}$ has a prior
distribution. Motivated by \cite{Raghavan10}, we assume a joint-geometric model for
the change propagation process where $\Gamma_{\pi_1}$ is geometric with parameter
$\rho$:
\begin{eqnarray} \label{eq:joint_geometric1}
\mathbb{P} \left(\Gamma_{\pi_1} = k \right) = \rho \cdot \left(1 - \rho \right)^{k-1},
\hspp
k = 1, 2, \cdots , \hspp \textrm{and} \hspp \mathbb{P} \left(\Gamma_{\pi_1} = 0 \right) = 0.
\end{eqnarray}
Further, $\Gamma_{\pi_{\ell}}$ is conditionally geometric with parameter
$\lambda_{\ell,\ell-1}$ for $\ell = 2, \cdots, L$:
\begin{align}
& \mathbb{P} \left(\Gamma_{\pi_{\ell }} = k+n \,|\, \Gamma_{\pi_{\ell - 1}} = n \right)
= \lambda_{\ell, \ell - 1} \cdot \left( 1 - \lambda_{\ell, \ell-1} \right)^k,
\hspp
k = 0, 1, 2, \cdots, \hspp \textrm{and} \hspp \ell = 2, \cdots, L.
\label{eq:joint_geometric2}
\end{align}
To simplify the analysis, we make the assumption that
$\lambda_{\ell, \ell -1} = \lambda$ for all $\ell = 2, \cdots, L$. Note that
$\rho \rightarrow 0$ corresponds to a ``rare'' change regime. On the other
hand, $\lambda \rightarrow 0$ corresponds to the case of uniformly likely
change propagation and $\lambda \rightarrow 1$ corresponds to the
case of instantaneous change propagation. In the extreme cases of $\lambda = 0$
and $\lambda = 1$, we can view the system as made of only one sensor and
$L$ sensors observing change at the same time-instant, respectively. We assume
that the parameters $\rho$ and $\lambda$ are known.

\subsection{Sensor Network Models}

We consider two different versions of the change detection problem in this paper:
centralized and decentralized. In both models, we assume that there are parallel communication
channels between sensor nodes and the fusion center.

\subsubsection{Centralized Model}

In the centralized setup, a central processor (fusion center) has
all the system-wide information. In particular, let
${\mathbf{z}}_k = \left[z_{k,1}, \hsppp z_{k,2}, \hsppp \cdots, \hsppp z_{k,L} \right]$
denote the collection of measurements at all sensors at time $k$. Then, the fusion center
has access to
\begin{gather} \nonumber
I_k = \{\mathbf{z}_1, \mathbf{z}_2, \dots, \mathbf{z}_k\},
\end{gather}
and  makes a decision based on $I_k$ at each time $k$.
Though sensors collect the measurements,
the fusion center is the only decision maker in the system. Hence, the centralized model is
equivalent to a single-sensor case with multi-dimensional vector of measurements at
each time. In practice, this model can be used for a sensor network located in a
geographically small area and when there is an ample bandwidth for reliable
communication between sensors and the fusion center.

\subsubsection{Decentralized Model}

In the decentralized setup, due to communication constraints, e.g., limited energy and
bandwidth, the fusion center has access to only a summary of the
measurements made at sensors and makes the decision based on local messages received from the sensors.
In particular, sensor ${\ell}$ produces a
local message $u_{k, \ell}$ at time $k$ from a finite-size alphabet\footnote{In
Section \ref{sec:event_based}, an event-triggered sampling scheme is used instead of
conventional uniform-in-time sampling.} $\{0, 1, \dots, U_{\ell} -1\}$.
Thus, both sensor nodes and the fusion center are involved in the decision-making process.

\begin{figure}
\center
  \includegraphics[width=100mm]{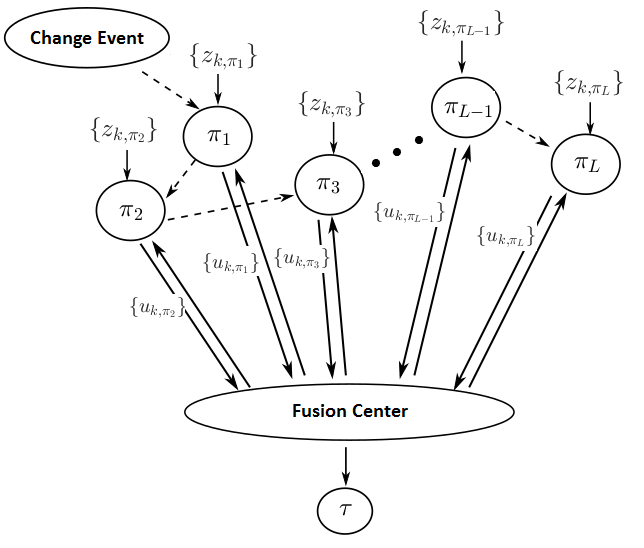}
  \vspace{-0.2cm}
\caption{The decentralized system model. Two-way communication between sensors and the fusion center is assumed where each sensor $\ell$ sends a message $u_{k,\ell}$ at time $k$ to the fusion center based on its measurement $z_{k,\ell}$. The fusion center then broadcasts all messages it has received to all sensors. After a change phenomenon occurs, it propagates through sensors as illustrated in the figure. Based on the received messages, the fusion center determines the stopping time $\tau$.}
 \label{fig:system_mod}
\end{figure}

Quasi-classical information structure enables the use of dynamic programming (DP) framework to obtain optimal solutions for sequential decision fusion problems and in particular decentralized sequential change detection problems \cite{Veeravalli99}. In this structure, all decision makers in the system must have the same past information. To satisfy this condition, we consider a system with full feedback and local memories. That is, two-way communication is employed where the sensor nodes send their local messages to the fusion center and the fusion center broadcasts the received messages to all sensors (see Fig. \ref{fig:system_mod}). Moreover, each sensor keeps in its local memory the local messages from all sensors in the previous time instants. In this way, all decision makers in the system have the same (one-time-instant delayed) past information.

Information available for decision-making at the fusion center at time $k$, i.e., $I_k$, is given as
\begin{gather} \label{eq:I_k} \nonumber
I_k = \{\mathbf{u}_1, \mathbf{u}_2, \dots, \mathbf{u}_k\},
\end{gather}
where $\mathbf{u}_k = [u_{k,1}, u_{k,2}, \dots, u_{k,L}]$ is the set of local messages at time $k$. The past information available at each sensor at time $k$ is $I_{k-1}$. Then, sensor $\ell$ quantizes $z_{k,\ell}$ through a local decision (or message) function $\phi_{k,\ell}$, i.e., $\phi_{k,\ell}(z_{k,\ell}) = u_{k,\ell}$ where $\phi_{k,\ell}$ depends on $I_{k-1}$. The set of all local decision functions at time $k$ is denoted with $\pmb{\phi}_k = [\phi_{k,1}, \dots, \phi_{k,L}]$.

\textit{Remark 2:} In practice, implementing a system using the described quasi-classical information structure is infeasible due to communication and storage constraints. In particular, two-way communications and broadcasting all the received messages to all sensors require significant communication resources which may not be available in a sensor network. These assumptions are just made to employ the DP arguments and to gain insights about the optimal solution structure. Later in this paper, we will derive asymptotically optimal local message functions and remove the need of feedback from the fusion center. Hence, the proposed final detection schemes will be based on one-way communications.

\subsection{Problem Formulation}

As stated before, change may propagate over sensor nodes in an arbitrary direction so the actual value of $\Pi$, i.e., the order of observing the change among the sensor nodes, is not known. We are particularly interested in detecting the first time any sensor observes a change, i.e., $\Gamma_{\pi_1}$. The Bayes risk is then expressed as
\begin{equation} \label{eq:Bayes_risk}
R(c) = \mathbb{P}(\tau < \Gamma_{\pi_1}) + c \, \mathbb{E} \, [(\tau - \Gamma_{\pi_1})^{+}],
\end{equation}
where $\tau$ is the stopping time determined by the fusion center, $\mathbb{P}(\tau < \Gamma_{\pi_1})$ is the probability of false alarm (PFA), $(\cdot)^+ = \max\{\cdot,0\}$, and $c \, \mathbb{E}[(\tau - \Gamma_{\pi_1})^{+}]$ corresponds to the average detection delay (ADD) where $c>0$ is a weight parameter indicating the cost of delay and the expectation is with respect to $\Gamma_{\pi_1}$.

There is a tradeoff between the performance metrics PFA and ADD whose values are determined based on the chosen stopping time in the fusion center and local message functions across sensor nodes. Our objective is to minimize the Bayes risk over all admissible stopping times $\tau$ and local quantizers $\{\pmb{\phi}_k\}_k$, which together form the policy $\Omega = \left\{\tau, \{\pmb{\phi}_k\}_k\right\}$. Note that in the centralized setting, the policy consists only of $\tau$, i.e., $\Omega = \tau$. Then, the optimization problem is stated as follows:
\begin{gather} \label{eq:opt_prob}
\mbox{minimize }  R(c) \mbox{  over all choices of } \Omega.
\end{gather}

We will employ the dynamic programming framework \cite{Bertsekas76} to derive the structure of the optimal solution to \eqref{eq:opt_prob}, as detailed in the following sections. Henceforth, we focus on the decentralized problem and meanwhile present the results for the centralized problem. In fact, the centralized problem can be considered as a special case of the decentralized problem where the feedback from the fusion center is removed, i.e., one-way communication only, and the local message functions are identity functions, i.e., $u_{k,\ell} = \phi_{k,\ell}(z_{k,\ell}) = z_{k,\ell}, ~ \forall \ell \in \{1, 2, \dots, L\}$ and $\forall k \geq 1$.

\section{Dynamic Programming Framework and The Finite-Horizon Case} \label{sec:dp}

\subsection{DP Formulation}

Firstly, we argue that the problem satisfies the Markov conditions for dynamic programming and hence can be solved under the DP framework. To this end, we define a discrete-time dynamic system with state $S_k$ representing the number of sensor nodes that have observed the change by time $k$, i.e.,
\begin{gather} \label{eq:state}
S_k = \sum_{\ell=1}^{L} \mathbb{I}(\Gamma_{\pi_\ell} \leq k),
\end{gather}
where $\mathbb{I}(\cdot)$ is an indicator function. Note that $S_0 = 0$ and $S_k \in \{0, 1, \dots, L\} \cup \Upsilon$ where $\Upsilon$ represents the terminal state. We assume that immediately after the fusion center declares the stopping time, the system moves into the terminal state and always stays there afterwards. Hence, the system state evolves over time as follows:
\begin{equation}\label{eq:state_evolution}
S_k =
\begin{cases}
  S_{k-1} + \sum_{\ell=1}^{L} \mathbb{I}(\Gamma_{\pi_\ell} = k), & \mbox{if } S_{k-1} \neq \Upsilon \mbox{ and } \tau \neq k, \\
  \Upsilon, & \mbox{if } S_{k-1} = \Upsilon \mbox{ or } \tau = k.
\end{cases}
\end{equation}

Note that \eqref{eq:state_evolution} can be written in a functional form as ${S_k = f(S_{k-1}, \tau, w_k)}$ where $w_k \triangleq \{\mathbb{I}(\Gamma_{\pi_\ell} = k),~ \ell=1, \dots, L\}$. Given $S_{k-1}$, it is clear that the next sensor that will observe the change after time $k-1$ is sensor $\pi_{S_k}$ and due to the defined joint-geometric process (cf. \eqref{eq:joint_geometric1} and \eqref{eq:joint_geometric2}), the probability of observing the change at sensor $\pi_{S_k}$ at time $k$ given $S_{k-1}$, i.e., $\mathbb{P}(\Gamma_{\pi_{S_k}} = k \, | \, S_{k-1})$ is equal to $\rho$ if $S_{k-1} = 0$, and $\lambda$, otherwise. Similarly, the probabilities of the sensors $\{\pi_\ell : \ell > S_k\}$ observing the change at time $k$ given $S_{k-1}$ can be simply calculated using \eqref{eq:joint_geometric1} and \eqref{eq:joint_geometric2}. Moreover, given $S_{k-1}$, it is clear that $\mathbb{I}(\Gamma_{\pi_\ell} = k) = 0$ for the sensors $\{\pi_\ell: \ell < S_k\}$ since they already observed the change before time $k$. Hence, given $S_{k-1}$, $w_k$ is conditionally independent from $w_0, w_1, \dots, w_{k-1}$, which implies that the state evolution equation satisfies the necessary Markov condition for dynamic programming.

It is not possible to observe the system state directly but inference about it can be done through sensor measurements. Using the state definition given \eqref{eq:state}, the measurement vector at time $k$ can be written in terms of the system state $S_k$ as
\begin{gather} \label{eq:obs_vector} \nonumber
\mathbf{z}_k = \mathbf{P}^{-1}_{\Pi} \begin{bmatrix} \mathbf{v}_{k}^{1} \\ \mathbf{v}_{k}^{0} \end{bmatrix} \mathbb{I}(S_k \neq \Upsilon),
\end{gather}
where $\mathbf{P}^{-1}_{\Pi}$ is inverse permutation matrix, $\mathbf{v}_{k}^{1} \in \mathbb{R}^{S_k}$ denotes the measurements with density $f_1$, i.e., $\{z_{k,\pi_\ell}: \Gamma_{\pi_\ell} \leq k\}$, and $\mathbf{v}_{k}^{0} \in \mathbb{R}^{(L - S_k)}$ denotes the measurements with density $f_0$, i.e., $\{z_{k,\pi_\ell}: \Gamma_{\pi_\ell} > k\}$. Note that when the system is in the terminal state, since the change decision is already made, sensors do not take measurements. Because the measurements are conditionally independent over time, the measurement vector also satisfies the necessary Markov condition for dynamic programming. Note that each row of a permutation matrix contains only one nonzero entry which is equal to 1 and the column of this entry is determined by $\Pi$. For example, if $\Pi = [3 \, 1 \, 2]$, then
\begin{equation}\nonumber
\mathbf{P}_{\Pi} = \left[
\begin{smallmatrix}
  0 & 0 & 1 \\
  1 & 0 & 0 \\
  0 & 1 & 0
\end{smallmatrix}\right].
\end{equation}

\subsection{Some Useful Results for the Finite-Horizon Case} \label{sec:finite_horizon}

Firstly, we restrict the stopping time $\tau$ to a finite-horizon $[0, T]$. Note that the original problem given in \eqref{eq:opt_prob} does not have such a restriction. Our purpose here is to provide some useful results that will later be helpful to characterize the solution in the infinite-horizon by letting $T \rightarrow \infty$. Using \eqref{eq:Bayes_risk} and the equality $(a-b)^+ = \sum_{k=0}^{a-1} \mathbb{I}(k \geq b)$ for integers $a, b \geq 0$, we obtain the following expression for the Bayes risk:
\begin{align} \label{eq:risk_v2_eq2} 
R(c)
&= \mathbb{P}(\tau < \Gamma_{\pi_1}) + c \, \mathbb{E} \bigg[\sum_{k=0}^{\tau-1} \mathbb{P}(k \geq \Gamma_{\pi_1})\bigg] \\ \label{eq:risk_v2}
&= \mathbb{P}(S_\tau = 0) + c \, \mathbb{E} \bigg[\sum_{k=0}^{\tau-1} \mathbb{P}(S_k \geq 1)\bigg],
\end{align}
where the expectations in \eqref{eq:risk_v2_eq2} and  \eqref{eq:risk_v2} are both with respect to $\Gamma_{\pi_1}$.
Note that \eqref{eq:risk_v2} is obtained from \eqref{eq:risk_v2_eq2} using the fact that the probability that sensor $\pi_1$ has observed the change by time $k$ is equal to the probability that the number of sensors that have observed the change by time $k$ is greater than or equal to $1$, i.e., $\mathbb{P}(k \geq \Gamma_{\pi_1}) =  \mathbb{P}(S_k \geq 1)$, which also implies that $\mathbb{P}(k < \Gamma_{\pi_1}) =  \mathbb{P}(S_k = 0)$.

The Bayes risk can be recursively minimized in the finite-horizon case using dynamic programming. Because the fusion center determines the stopping rule $\tau$ and the information available at the fusion center at time $k$ is $I_k$, the minimum expected cost-to-go at time $k$ is a function of $I_k$ and denoted with $\tilde{J}_{k}^{T}(I_{k})$. Then, according to the Bayes risk given in \eqref{eq:risk_v2}, the finite-horizon DP equations are given as (see \cite{Bertsekas76} for similar examples)
\begin{gather} \nonumber
\tilde{J}_{T}^{T}(I_T) = \mathbb{P}(S_T = 0 \,|\, I_T),
\end{gather}
and for $0 < k \leq T$,
\begin{gather} \label{eq:finiteDPk_v1}
\tilde{J}_{k-1}^{T}(I_{k-1}) = \min\bigg\{ \mathbb{P}(S_{k-1} = 0 \,|\, I_{k-1}), \, c \, \mathbb{P}(S_{k-1} \geq 1 \,|\, I_{k-1}) + \min_{\pmb{\phi}_k} \big\{ {\mathbb{E}\,[\tilde{J}_{k}^{T}(I_{k})\,|\,I_{k-1}]} \big\} \, \bigg\}.
\end{gather}
In \eqref{eq:finiteDPk_v1}, the first term in the outer minimum corresponds to the cost of stopping and declaring a change at time $k-1$ and the other term corresponds to expected cost of proceeding to time $k$. Note that the minimum Bayes risk is equal to $\tilde{J}_{0}^{T}(I_0)$ where $I_0$ is an empty set.

We observe through \eqref{eq:finiteDPk_v1} that the local message functions at time $k$, i.e., $\pmb{\phi}_k$, are chosen to minimize $\mathbb{E}\,[\tilde{J}_{k}^{T}(I_{k})\,|\,I_{k-1}]$, i.e., the expected cost-to-go at time $k$ given the information available at each sensor at time $k$, i.e., $I_{k-1}$. In the following, we firstly determine the structure of the optimal $\pmb{\phi}_k$ in the finite-horizon case. We then give a sufficient statistic for dynamic programming and derive its recursion over time. Finally, we express the finite-horizon DP equations in terms of this sufficient statistic and derive some properties related to the finite-horizon DP equations.

The following proposition states that at each time, for each sensor, quantizing the likelihood ratio of its measurement and transmitting the corresponding finite bit sequence to the fusion center is optimal where the quantization thresholds at time $k$ depend on $I_{k-1}$.

\textbf{Proposition 1:} Given a finite alphabet size for sensor messages, likelihood ratio (LR) quantizers with thresholds depending on $I_{k-1}$ are the optimal finite-horizon local message functions at time $k$. That is, for each sensor $\ell$, if the alphabet size is $U_\ell$, there are $U_\ell+1$ quantization thresholds, as a function of $I_{k-1}$ at time $k$, defining the $U_\ell$ quantization intervals. Let $\varphi_{k,\ell,1}$ and $\varphi_{k,\ell,2}$ be two such consecutive quantization thresholds defining an interval $(\varphi_{k,\ell,1},\varphi_{k,\ell,2}]$. If
\begin{gather} \nonumber
\varphi_{k,\ell,1} < L(z_{k,\ell}) \leq \varphi_{k,\ell,2},
\end{gather}
then $u_{k,\ell} = u$  where
\begin{gather} \nonumber
L(z_{k,\ell}) \triangleq \frac{{f_1}(z_{k,\ell})}{{f_0}(z_{k,\ell})}
\end{gather}
is the LR for $z_{k,\ell}$ and $u \in \{0, 1, \dots, U_{\ell}-1\}$.
\begin{proof}
See Appendix \ref{sec:proof_thm1}.
\end{proof}

Next, based on the DP equations given in \eqref{eq:finiteDPk_v1}, we define an $(L+1)$-dimensional vector of conditional probabilities $\mathbf{q}_k \triangleq [q_{k}^{0}, q_{k}^{1}, \dots, q_{k}^{L}]$ where $q_{k}^{n} \triangleq \mathbb{P}(S_k = n\,|\,I_k)$. Our aim is to show that $\mathbf{q}_k$ is a sufficient statistic for dynamic programming and it can be recursively updated over time. We can write
\begin{align} \nonumber
q_{k}^{n} &= \mathbb{P}(S_k = n\,|\,I_k) = \mathbb{P}(S_k = n\,|\,\mathbf{u}_k, I_{k-1}) \\ \nonumber
&= \frac{\mathbb{P}(\mathbf{u}_k \,|\, S_k = n,\,I_{k-1}) \mathbb{P}(S_k = n\,|\,I_{k-1})}{\sum_{m=0}^{L} \mathbb{P}(\mathbf{u}_k\,|\,S_k = m,\,I_{k-1}) \mathbb{P}(S_k = m \,|\, I_{k-1})} \\ \label{recursion}
&\triangleq \frac{g_{k}^{n}}{\sum_{m=0}^{L} g_{k}^{m}},
\end{align}
where
\begin{gather} \label{eq:gkn_tmp}
g_k^n \triangleq \mathbb{P}(\mathbf{u}_k \,|\, S_k = n,\,I_{k-1}) \, \mathbb{P}(S_k = n\,|\,I_{k-1}).
\end{gather}

We would like to express $g_k^n$ in terms of $\mathbf{q}_{k-1}$. To this end, we write each probability on the RHS of \eqref{eq:gkn_tmp} more explicitly as follows:
\begin{align} \nonumber
\mathbb{P}(\mathbf{u}_k \,|\, S_k = n,\,I_{k-1})
&= \sum_{\Pi} \underbrace{\mathbb{P}(\Pi = \Pi_j)}_{1/L!} \, \mathbb{P}(\mathbf{u}_k \,|\, \Pi = \Pi_j,\, S_k = n,\,I_{k-1}) \\ \nonumber
&= \frac{1}{L!} \, \sum_{\Pi} \bigg( \, \prod_{i=1}^{n} \mathbb{P}(u_{k,\pi_{i}} \,|\, \Gamma_{\pi_{i}} \leq k, \,I_{k-1})
 \prod_{i=n+1}^{L} \mathbb{P}(u_{k,\pi_{i}} \,|\, \Gamma_{\pi_{i}} > k, \,I_{k-1}) \, \bigg) \\ \label{eq:tmpp1}
&= \frac{1}{L!} \, \sum_{\Pi} \bigg( \, \prod_{i=1}^{n} \mathbb{P}_1(\phi_{k,\pi_{i}}(z_{\pi_{i}}) = u_{k,\pi_{i}})
 \prod_{i=n+1}^{L} \mathbb{P}_0(\phi_{k,\pi_{i}}(z_{\pi_{i}}) = u_{k,\pi_{i}}) \, \bigg),
\end{align}
where $\Pi_j$ denotes a pattern among the $L!$ possible patterns and $z_{\pi_{i}}$ denotes a generic observation at sensor ${\pi_{i}}$. Moreover, $\mathbb{P}_1(\cdot) \triangleq \mathbb{P}(\cdot \, | \, H_1)$ and $\mathbb{P}_0(\cdot) \triangleq \mathbb{P}(\cdot \, | \, H_0)$ denote the probabilities under hypotheses $H_1$ and $H_0$, respectively where $H_i$ corresponds to the case where the pdf of measurements is $f_i$, $i \in \{0,1\}$. Furthermore,
\begin{align}\nonumber
\mathbb{P}(S_k = n\,|\,I_{k-1})
&= \sum_{m=0}^{n} \underbrace{\mathbb{P}(S_{k-1} = m\,|\,I_{k-1})}_{q_{k-1}^{m}} \, \mathbb{P}(S_k = n\,|\, S_{k-1} = m,\, I_{k-1}) \\ \nonumber
&= \begin{cases}
    q_{k-1}^{0} \, (1 - \rho), & \mbox{if } n = 0, \\
    \big(q_{k-1}^{0} \, \rho \, \lambda^{n-1} + \sum_{m=1}^{n} q_{k-1}^{m} \, \lambda^{n-m}\big) (1-\lambda), & \mbox{if } n \in \{1, 2, \dots, L-1\}, \\
    q_{k-1}^{0} \, \rho \, \lambda^{L-1} + \sum_{m=1}^{L} q_{k-1}^{m} \, \lambda^{L-m}, & \mbox{if } n = L,
  \end{cases} \\ \label{eq:tmpp2}
&= (1-\rho_{\pi_{n},\pi_{n+1}}) \sum_{m=0}^{n} q_{k-1}^{m} \, e_{m}^{n},
\end{align}
where we use the following notations:
\begin{gather} \nonumber
e_{m}^{n} \triangleq \prod_{\ell=m}^{n-1} \rho_{\pi_{\ell},\pi_{\ell+1}},
\end{gather}
and
\begin{equation} \label{eq:rhos} \nonumber
    \rho_{\pi_{\ell},\pi_{\ell+1}} \triangleq
    \begin{cases}
     \rho , & \text{if} ~~ \ell = 0, \\
     \lambda , & \text{if} ~~ \ell \in \{1, 2, \dots, L-1\}, \\
     0 , & \text{if} ~~ \ell = L,
    \end{cases}
\end{equation}
with $e_{m}^{m} = 1$ for any ${m \in \{0, 1, \dots, L\}}$.

Substituting \eqref{eq:tmpp1} and \eqref{eq:tmpp2} into \eqref{eq:gkn_tmp}, we have
\begin{gather} \label{eq:recursion_g_v2}
g_k^n = \frac{1}{L!} \, \sum_{\Pi} \bigg( \, \prod_{i=1}^{n} \mathbb{P}_{1}(\phi_{k,\pi_{i}}(z_{\pi_{i}}) = u_{k,\pi_{i}})
 \prod_{i=n+1}^{L} \mathbb{P}_{0}(\phi_{k,\pi_{i}}(z_{\pi_{i}}) = u_{k,\pi_{i}}) \, \bigg)
 \, \bigg( \, (1-\rho_{\pi_{n},\pi_{n+1}}) \sum_{m=0}^{n} q_{k-1}^{m} \, e_{m}^{n}  \, \bigg).
 \end{gather}
Since $g_{k}^{n}$ is a function of $\mathbf{u}_k$, $\pmb{\phi}_{k}$, and $\mathbf{q}_{k-1}$, we denote it with $g_{k}^{n}(\mathbf{u}_k, \pmb{\phi}_{k}, \mathbf{q}_{k-1})$ and define
\begin{gather} \label{eq:p_uk_ik_1}
h_k(\mathbf{u}_k, \pmb{\phi}_{k}, \mathbf{q}_{k-1}) \triangleq \mathbb{P}(\mathbf{u}_k\,|\,I_{k-1}) = \sum_{m=0}^{L} g_{k}^{m}(\mathbf{u}_k, \pmb{\phi}_{k}, \mathbf{q}_{k-1}).
\end{gather}
Then from \eqref{recursion}, we have
\begin{gather} \label{eq:recursion_v2}
q_{k}^{n} = \frac{g_{k}^{n}(\mathbf{u}_k, \pmb{\phi}_{k}, \mathbf{q}_{k-1})}{h_k(\mathbf{u}_k, \pmb{\phi}_{k}, \mathbf{q}_{k-1})}.
\end{gather}
Here, we note that $q_{k}^{n}$ depends on $\mathbf{q}_{k-1}$, not just $q_{k-1}^{n}$. This is the reason for considering the entire vector $\mathbf{q}_k$ as a sufficient statistic at time $k$. Because the RHS of \eqref{eq:recursion_v2} includes $\pmb{\phi}_{k}$ which depends on $I_{k-1}$ as stated in Proposition 1, in order to make  \eqref{eq:recursion_v2} a useful recursive formula, we need an additional result that $\pmb{\phi}_{k}$ can be expressed as a function of only $\mathbf{q}_{k-1}$. For this purpose, we present Lemma 1 below. 

\textbf{Lemma 1:} (a) $\tilde{J}_{k}^{T}(I_k)$ can be written as a function of only $\mathbf{q}_k$ for each $k$, $0 \leq k \leq T$. Let this function be denoted with $J_{k}^{T}(\mathbf{q}_k)$. \\
(b) The optimal $\pmb{\phi}_k$ can be written as a function of only $\mathbf{q}_{k-1}$ for ${0 < k \leq T}$.

\begin{proof}
The proof is similar to \cite[Proof of Proposition 3]{Veeravalli93}. Firstly, we observe that $\tilde{J}_{T}^{T}(I_T) = \mathbb{P}(S_T = 0 \,|\, I_T) = q_T^0$. Hence, for $k=T$, clearly we can write $J_{T}^{T}(\mathbf{q}_T) = q_T^0$. We now use an induction argument. Suppose that for any $k \in \{1,2,\dots,T-1 \}$, $\tilde{J}_{k}^{T}(I_k)$ can be written as a function of only $\mathbf{q}_k$, which is denoted with $J_{k}^{T}(\mathbf{q}_k)$. Then, using \eqref{eq:finiteDPk_v1}, \eqref{eq:p_uk_ik_1}, and \eqref{eq:recursion_v2}, we have
\begin{align} \nonumber
\tilde{J}_{k-1}^{T}(I_{k-1})
&= \min\bigg\{ q_{k-1}^0, \, c \, (1 - q_{k-1}^0) + \min_{\pmb{\phi}_k} \big\{ {\mathbb{E}\,[J_{k}^{T}(\mathbf{q}_k)\,|\,I_{k-1}]} \big\} \, \bigg\} \\ \label{eq:lemma1_proof}
&= \min\bigg\{ q_{k-1}^0, \, c \, (1 - q_{k-1}^0) + \min_{\pmb{\phi}_k} \big\{ \sum_{\mathbf{u}_k} J_{k}^{T}\bigg(\frac{\mathbf{g}_{k}(\mathbf{u}_k, \pmb{\phi}_{k}, \mathbf{q}_{k-1})}{h_k(\mathbf{u}_k, \pmb{\phi}_{k}, \mathbf{q}_{k-1})}\bigg) {h_k(\mathbf{u}_k, \pmb{\phi}_{k}, \mathbf{q}_{k-1})} \big\} \, \bigg\},
\end{align}
where
\begin{gather} \nonumber
\mathbf{g}_k(\cdot) \triangleq \big[g_{k}^{0}(\cdot), g_{k}^{1}(\cdot), \dots, g_{k}^{L}(\cdot)\big].
\end{gather}

In \eqref{eq:lemma1_proof}, we observe that the optimal ${\pmb{\phi}_k}$ is chosen to minimize a function depending on ${\pmb{\phi}_k}$ and $\mathbf{q}_{k-1}$. Hence, the optimal ${\pmb{\phi}_k}$ depends only on $\mathbf{q}_{k-1}$. Through \eqref{eq:lemma1_proof}, we also observe that $\tilde{J}_{k-1}^{T}(I_{k-1})$ is a function depending only on $\pmb{\phi}_{k}$ and $\mathbf{q}_{k-1}$. Since the optimal $\pmb{\phi}_{k}$ is a function of only $\mathbf{q}_{k-1}$, $\tilde{J}_{k-1}^{T}(I_{k-1})$ can be written as a function of only $\mathbf{q}_{k-1}$, which is denoted with $J_{k-1}^{T}(\mathbf{q}_{k-1})$. This completes the proofs of (a) and (b).
\end{proof}

By Lemma 1-(b), \eqref{eq:recursion_v2} gives the recursive formula to obtain $q_k^n$ from $\mathbf{q}_{k-1}$ at time $k$ if the optimal local quantizers are used. Then, using \eqref{eq:recursion_v2} for all $n \in \{1,\dots,L\}$, we obtain the recursion of sufficient statistics:
\begin{gather} \label{eq:recursion_v7}
\mathbf{q}_{k} = \frac{\mathbf{g}_{k}(\mathbf{u}_k, \pmb{\phi}_{k}, \mathbf{q}_{k-1})}{h_k(\mathbf{u}_k, \pmb{\phi}_{k}, \mathbf{q}_{k-1})}.
\end{gather}

Lemma 1-(a) states that the finite-horizon cost-to-go function at each time $k$ can be expressed as a function of $\mathbf{q}_k$. Note that $\sum_{n=0}^{L} q_{k}^{n} = 1$ and $\mathbb{P}(S_{k} = 0 \,|\, I_{k}) = q_{k}^{0}$. Then, $\mathbb{P}(S_{k} \geq 1 \,|\, I_{k}) = 1 - q_{k}^{0}$. Using Lemma 1-(a) and these equalities, we write the finite-horizon DP equations given in \eqref{eq:finiteDPk_v1} in terms of sufficient statistics as follows:
\begin{gather} \nonumber
J_{T}^{T}(\mathbf{q}_T) = q_{T}^{0},
\end{gather}
and for $0 < k \leq T$,
\begin{gather} \label{eq:finiteDPk_v2}
J_{k-1}^{T}(\mathbf{q}_{k-1}) = \min\big\{ q_{k-1}^{0}, c \, (1 - q_{k-1}^{0}) + A_{k-1}^{T} (\mathbf{q}_{k-1}) \big\},
\end{gather}
where
\begin{align} \nonumber
A_{k-1}^{T}(\mathbf{q}_{k-1}) &= \min_{\pmb{\phi}_{k}} {\mathbb{E}\,[J_{k}^{T}(\mathbf{q}_k)\,|\,I_{k-1}]} \\ \label{A_k_T_v2}
 &= \min_{\pmb{\phi}_{k}} \bigg\{\sum_{\mathbf{u}_{k}} J_{k}^{T}\bigg(\frac{\mathbf{g}_k(\mathbf{u}_k, \pmb{\phi}_{k}, \mathbf{q}_{k-1})}{h_k(\mathbf{u}_{k}, \pmb{\phi}_{k}, \mathbf{q}_{k-1})}\bigg) h_k(\mathbf{u}_{k}, \pmb{\phi}_{k}, \mathbf{q}_{k-1}) \bigg\}.
\end{align}

Using the structure of DP equations given above, we can derive some properties of the functions $J_{k}^{T}$ and $A_{k}^{T}$ that can be useful in characterizing the solution structure for the infinite-horizon case. The following lemma states these properties.

\textbf{Lemma 2:}
(a) $J_{k}^{T}(\mathbf{q})$ and $A_{k}^{T}(\mathbf{q})$ are concave functions of $\mathbf{q}$ over the $L$-dimensional simplex $\mathcal{Q} \triangleq \{\mathbf{q}: \sum_{n=0}^{L} q^{n} = 1, q^{n} \geq 0, \forall n \in \{0,1,\dots,L\}\}$. \\
(b) $0 \leq J_{k}^{T}(\mathbf{q}), A_{k}^{T}(\mathbf{q}) \leq 1, \forall \mathbf{q} \in \mathcal{Q}$ and $J_{k}^{T}(\mathbf{q}) = A_{k}^{T}(\mathbf{q}) = 0$ over the hyperplane $\mathcal{P} \triangleq {\{\mathbf{q}: q^{0} = 0\}}$.

\begin{proof}
See Appendix \ref{sec:proof_lemma2}.
\end{proof}

Before studying the infinite-horizon case, we finally note that based on Proposition 1 and Lemma 1-(b), the optimal $\pmb{\phi}_k$ are LR quantizers with thresholds depending on $\mathbf{q}_{k-1}$. As stated before, the LR quantizer maps an interval of LRs to a number in a finite alphabet and the optimal $\pmb{\phi}_k$ at each time $k$ are chosen to minimize ${\mathbb{E}\,[J_{k}^{T}(\mathbf{q}_k)\,|\,I_{k-1}]}$. In \eqref{A_k_T_v2}, we observe that ${\mathbb{E}\,[J_{k}^{T}(\mathbf{q}_k)\,|\,I_{k-1}]}$ depends on $\pmb{\phi}_k$ only through the functions $\mathbf{g}_k$ and $h_k$ and in \eqref{eq:recursion_g_v2}, we notice that only the likelihoods of the LR intervals are needed to calculate these functions. Therefore, the output of an LR quantizer is related only to the input quantization interval. Then, using a permutation mapping, the LR quantization outputs can be made increasing with the quantization thresholds without changing the value of ${\mathbb{E}\,[J_{k}^{T}(\mathbf{q}_k)\,|\,I_{k-1}]}$ at each time $k$. Hence, $\pmb{\phi}_{k}$ can be restricted to the class of monotone LR quantizers with thresholds depending on $\mathbf{q}_{k-1}$ without loss of optimality. That is, at time $k$, for a sensor, say $\ell$, there exists a set of thresholds
\begin{gather} \nonumber
0 = \varphi_{k,\ell,0} \leq \varphi_{k,\ell,1} \leq \dots \leq \varphi_{k,\ell,U_{\ell-1}} \leq \varphi_{k,\ell,U_\ell} = \infty
\end{gather}
depending on $\mathbf{q}_{k-1}$ such that $u_{k,\ell} = i$ only if
\begin{gather} \nonumber
\varphi_{k,\ell,i} < L(z_{k,\ell}) \leq \varphi_{k,\ell,{i+1}}, ~~ i = 0, 1, \dots, U_{\ell-1}.
\end{gather}
Note that since the LR is always nonnegative, we have $\varphi_{k,\ell,0} = 0$ and $\varphi_{k,\ell,U_\ell} = \infty, \forall \ell \in \{1,2, \dots, L\}$ and $\forall k \geq 1$. Let $\pmb{\Phi}$ be the set of monotone LR quantizers. Then, $\pmb{\phi}_{k} \subset \pmb{\Phi}$. Hence, considering all sensors, determining the optimal quantizers requires to optimize $\sum_{\ell=1}^{L} (U_{\ell}-1)$ thresholds at each time $k \geq 1$.

\section{Optimal Solution Structure} \label{sec:inf_horizon}

Now, we remove the restriction on the stopping time $\tau$ by letting $T \rightarrow \infty$. The following proposition will be useful in characterizing the structure of the optimal stopping rule.

\textbf{Proposition 2:} In the infinite-horizon case, we have the following Bellman equation:
\begin{gather} \label{eq:bellman}
J(\mathbf{q}) = \min\{q^{0}, \, c \, (1 - q^{0}) + A(\mathbf{q})\},
\end{gather}
where
\begin{gather} \nonumber
J(\mathbf{q}) \triangleq \lim_{T \rightarrow \infty}{J_{k}^{T}(\mathbf{q})} 
\end{gather}
is the infinite-horizon cost-to-go function and
\begin{align} \nonumber
A(\mathbf{q}) &\triangleq \lim_{T \rightarrow \infty} A_{k}^{T} (\mathbf{q}) \\ \label{eq:A_q}
&= \min_{\pmb{\phi} \, \subset \, \pmb{\Phi}} \bigg\{\sum_{\mathbf{u}} J\bigg(\frac{\mathbf{g}(\mathbf{u}, \pmb{\phi}, \mathbf{q})}{h(\mathbf{u}, \pmb{\phi}, \mathbf{q})}\bigg) h(\mathbf{u}, \pmb{\phi}, \mathbf{q}) \bigg\}.
\end{align}
Furthermore, $J(\mathbf{q})$ and $A(\mathbf{q})$ are concave functions of $\mathbf{q}$ over the simplex $\mathcal{Q}$. Moreover, $0 \leq J(\mathbf{q}), A(\mathbf{q}) \leq 1, \forall \mathbf{q} \in \mathcal{Q}$ and $J(\mathbf{q}) = A(\mathbf{q}) = 0$ over the hyperplane $\mathcal{P}$.
\begin{proof}
First, we show that $\lim_{T \rightarrow \infty}{J_{k}^{T}(\mathbf{q})}$ exists. From \eqref{eq:finiteDPk_v2}, we have $J_{T}^{T+1}(\mathbf{q}_T) \leq q_{T}^{0} = J_{T}^{T}(\mathbf{q}_T)$ that implies $\mathbb{E} \, [J_{T}^{T+1}(\mathbf{q}_T)] \leq \mathbb{E} \, [J_{T}^{T}(\mathbf{q}_T)]$. Then, using \eqref{A_k_T_v2}, we have  $A_{T-1}^{T+1}(\mathbf{q}_{T-1}) \leq A_{T-1}^{T}(\mathbf{q}_{T-1})$, which using \eqref{eq:finiteDPk_v2} implies that $J_{T-1}^{T+1}(\mathbf{q}_{T-1}) \leq J_{T-1}^{T}(\mathbf{q}_{T-1})$. Continuing like this, it can be easily shown that ${J_{k}^{T+1}(\mathbf{q})} \leq {J_{k}^{T}(\mathbf{q})}, \forall k \in \{0,1,\dots,T\}$. Moreover, since ${J_{k}^{T}(\mathbf{q})} \geq 0, \forall \mathbf{q} \in \mathcal{Q}, \forall k \in \{0, 1, \dots, T\}$, and $\forall T \geq k$, $\lim_{T \rightarrow \infty}{J_{k}^{T}(\mathbf{q})} = {J_{k}^{\infty}(\mathbf{q})} = \inf_{T: \, T>k} {J_{k}^{T}(\mathbf{q})}$ exists. Further, since the defined joint-geometric process is memoryless and the measurements are i.i.d. in time, the starting point in time, i.e., $k$, is not important in determining the infinite-horizon cost-to-go function, i.e., ${J_{k+1}^{\infty}(\mathbf{q})} = {J_{k}^{\infty}(\mathbf{q})}$ for any $k \geq 0$. We then denote this limit as $J(\mathbf{q}) \triangleq \lim_{T \rightarrow \infty}{J_{k}^{T}(\mathbf{q})}$. Furthermore,
\begin{align} \label{eq:drrnt}
  \lim_{T \rightarrow \infty} A_{k}^{T} (\mathbf{q})
  &= \lim_{T \rightarrow \infty} \min_{\pmb{\phi} \, \subset \, \pmb{\Phi}} \bigg\{ \mathbb{E}\,\bigg[J_{k+1}^{T}\bigg(\frac{\mathbf{g}(\mathbf{u}, \pmb{\phi}, \mathbf{q})}{h(\mathbf{u}, \pmb{\phi}, \mathbf{q})}\bigg)\,\bigg{|}\,I_{k}\bigg] \bigg\} \\ \label{eq:drrnt22}
  &= \min_{\pmb{\phi} \, \subset \, \pmb{\Phi}} \bigg\{\sum_{\mathbf{u}} J\bigg(\frac{\mathbf{g}(\mathbf{u}, \pmb{\phi}, \mathbf{q})}{h(\mathbf{u}, \pmb{\phi}, \mathbf{q})}\bigg) h(\mathbf{u}, \pmb{\phi}, \mathbf{q}) \bigg\},
\end{align}
which follows due to the dominated convergence theorem \cite{Durrett10}, i.e., since by Lemma 2-(b), $J_{k}^{T}(\mathbf{q}) \leq 1, \forall k \in \{0,1,\dots,T\}, \forall \mathbf{q} \in \mathcal{Q}$, the limit operator can be written inside the expectation operator in \eqref{eq:drrnt} and since $\lim_{T \rightarrow \infty}{J_{k+1}^{T}(\mathbf{q})} = J(\mathbf{q}), \forall \mathbf{q} \in \mathcal{Q}$, \eqref{eq:drrnt22} is obtained. Hence, $\lim_{T \rightarrow \infty} A_{k}^{T} (\mathbf{q}_k)$ also exists and is denoted with $A(\mathbf{q})$. Finally, the stated properties of the functions $J$ and $A$ in the proposition are directly obtained using Lemma 2 and by letting $T \rightarrow \infty$.
\end{proof}

In \eqref{eq:bellman}, the infinite-horizon cost-to-go function is given where at each time $k$, $q_k^{0}$ is the cost of stopping and $c \, (1 - q_k^{0}) + A(\mathbf{q}_k)$ is the expected cost of continuing. The optimal stopping time is the first time at which the cost of stopping is smaller than or equal to the expected cost of continuing \cite[Sec. 7.2]{Bertsekas76}, i.e.,
\begin{align} \nonumber
\tau &= \min\{k \in \mathbb{N}: J(\mathbf{q}_k) = q_{k}^{0} \} \\ \label{eq:inf_horizon_sol_v0}
&= \min\{k \in \mathbb{N}: q_{k}^{0} \leq c \, (1 - q_k^{0}) + A(\mathbf{q}_k) \}.
\end{align}
Hence, the change is declared at the first time the concave function $c \,(1 - q^{0}) + A(\mathbf{q})$ exceeds the affine function $q^{0}$ of the sufficient statistics $\mathbf{q}$. Note that the minimum Bayes risk in the infinite-horizon is equal to $J(\mathbf{q}_0)$ where $\mathbf{q}_0$ is the initial value of the sufficient statistic at time $k=0$. Since $S_0 = 0$, we have $q_0^0 = 1$ and $q_0^n = 0, \forall n \in \{1, 2, \dots, L\}$.

Although \eqref{eq:inf_horizon_sol_v0} specify the optimal stopping rule, (i) having further characterization on it is difficult (except in a special case where the vector of sufficient statistics can be reduced to a one-dimensional vector, which is possible in the cases where $\lambda = 1$ \cite{Veeravalli01} and $\lambda = 0$), (ii) analyzing the performance metrics ADD and PFA of the optimal scheme is difficult, and (iii) it does not present a useful method to determine the optimal local message functions. Nevertheless, the following proposition leads to a numerical method to approximately compute the optimal solution and presents the structure of the optimal local message functions in the infinite-horizon case.

\textbf{Proposition 3:} (a) Let $\mathcal{S}$ be the set of all concave functions upper bounded by the function $f(\mathbf{q}) = q^{0}$ over the simplex $\mathcal{Q}$. Then, for any $B \in \mathcal{S}$, we define a function $K_B$ as follows:
\begin{gather} \label{eq:K_B}
K_B(\pmb{\phi}, \mathbf{q}) \triangleq  \sum_{\mathbf{u}} B\bigg(\frac{\mathbf{g}(\mathbf{u}, \pmb{\phi}, \mathbf{q})}{h(\mathbf{u}, \pmb{\phi}, \mathbf{q})}\bigg) h(\mathbf{u}, \pmb{\phi}, \mathbf{q}).
\end{gather}
Further, we define a mapping $\omega: \mathcal{S}\rightarrow\mathcal{S}$ over any function $B \in \mathcal{S}$ as follows:
\begin{gather} \label{eq:mapp}
\omega \, B(\mathbf{q}) \triangleq \min\{q^{0}, \, c \, (1 - q^{0}) + \min_{\pmb{\phi} \, \subset \, \pmb{\Phi}} K_B(\pmb{\phi}, \mathbf{q})\},
\end{gather}
i.e., $\omega$ maps a function $B(\mathbf{q}), \mathbf{q} \in \mathcal{Q}$ to another function $\omega \, B(\mathbf{q}), \mathbf{q} \in \mathcal{Q}$. Then, $J(\mathbf{q})$ given in \eqref{eq:bellman} is the unique fixed point of the mapping $\omega$.

(b) The infinite-horizon optimal local message functions are stationary in time, i.e., the optimal $\pmb{\phi}_k$ are the same for all $k \geq 1$.

\begin{proof}
See Appendix \ref{sec:proof_prop1}.
\end{proof}

According to Proposition 3-(b), the optimal local message functions are monotone LR quantizers with stationary threshold functions, which depend on the sufficient statistics. Let the set of optimal threshold functions be denoted with $\pmb{\varphi}_{\text{opt}} \triangleq [\pmb{\varphi}_{\text{opt},1}, \dots, \pmb{\varphi}_{\text{opt},L}]$ where $\pmb{\varphi}_{\text{opt},\ell} \triangleq [\varphi_{\ell,0}, \varphi_{\ell,1}, \dots, \varphi_{\ell,U_\ell-1}, \varphi_{\ell,U_\ell}]$ represents the thresholds as a function of sufficient statistics for sensor $\ell$. Hence, at each time $k$, the set of thresholds can be expressed as $\pmb{\varphi}_k \triangleq [\pmb{\varphi}_{k,1}, \dots, \pmb{\varphi}_{k,L}] \triangleq \pmb{\varphi}_{\text{opt}}(\mathbf{q}_{k-1})$ where $\pmb{\varphi}_{k,\ell} \triangleq [\varphi_{k,\ell,0}, \varphi_{k,\ell,1}, \dots, \varphi_{k,\ell,U_\ell}]$ are the thresholds at sensor $\ell$ at time $k$. Because $\varphi_{\ell,0} = 0$ and $\varphi_{\ell,U_\ell} = \infty$, ${\forall \ell \in \{1, \dots, L\}}$, all local message functions can be defined with ${\sum_{\ell=1}^{L} (U_\ell-1)}$ threshold functions and hence $\pmb{\phi}$ and $\pmb{\varphi}$ can be interchangeably used.

Then, $g_{k}^{n}(\cdot)$ in \eqref{eq:recursion_g_v2} can be rewritten as
\begin{gather} \nonumber
g_{k}^{n}(\mathbf{u}_k, \pmb{\varphi}_k, \mathbf{q}_{k-1}) = \frac{1}{L!} \, \sum_{\Pi} \bigg( \,
 \, \prod_{i=1}^{n} \mathbb{P}_1(\varphi_{k,\pi_{i}, u_{k,\pi_{i}}} < L(z_{\pi_{i}}) \leq \varphi_{k,\pi_{i},(u_{k,\pi_{i}}+1)}) \\ \label{eq:g_inf}
\times \, \prod_{i=n+1}^{L} \mathbb{P}_0(\varphi_{k,\pi_{i}, u_{k,\pi_{i}}} < L(z_{\pi_{i}}) \leq \varphi_{k,\pi_{i},(u_{k,\pi_{i}}+1)}) \, \bigg)
 \, \bigg( \, (1-\rho_{\pi_{n},\pi_{n+1}}) \sum_{m=0}^{n} q_{k-1}^{m} \, e_{m}^{n}  \, \bigg),
\end{gather}
where $L(z_{\pi_{i}})$ is the LR for a generic observation $z_{\pi_{i}}$ at sensor ${\pi_{i}}$. Moreover, $\varphi_{k,\pi_{i}, u_{k,\pi_{i}}} \triangleq \varphi_{\pi_{i}, u_{k,\pi_{i}}}(\mathbf{q}_{k-1})$ and $\varphi_{k,\pi_{i}, (u_{k,\pi_{i}}+1)} \triangleq \varphi_{\pi_{i}, (u_{k,\pi_{i}}+1)}(\mathbf{q}_{k-1})$ denote the thresholds defining the LR quantization interval $(\varphi_{k,\pi_{i}, u_{k,\pi_{i}}}, \varphi_{k,\pi_{i}, (u_{k,\pi_{i}}+1)}]$ at sensor ${\pi_{i}}$ at time $k$ when the local message received from sensor ${\pi_{i}}$ is $u_{k,\pi_{i}} \in \{0,1,\dots,U_{\ell-1}\}$. Further, the mapping $\omega$ can be rewritten for any $B \in \mathcal{S}$ as
\begin{gather} \label{eq:mapp_inf} \nonumber
\omega \, B(\mathbf{q}) = \min\{q^{0}, \, c \, (1 - q^{0}) + \min_{\pmb{\varphi}} K_B(\pmb{\varphi}, \mathbf{q})\}.
\end{gather}

Then, for each $\mathbf{q} \in \mathcal{Q}$, clearly we have
\begin{gather} \nonumber
\min\{q^{0}, \, c \, (1 - q^{0}) + \min_{\pmb{\varphi}} K_f(\pmb{\varphi}, \mathbf{q})\} \leq q^{0} \triangleq f(\mathbf{q}),
\end{gather}
which shows that $\omega f(\mathbf{q}) \leq f(\mathbf{q}), \forall \mathbf{q} \in \mathcal{Q}$. We now use induction to show $\omega^{i+1} f(\mathbf{q}) \leq \omega^i f(\mathbf{q}), \forall \mathbf{q} \in \mathcal{Q}$ and for any nonnegative integer $i$. Suppose that for any integer $i > 2$,
$\omega^{i} f(\mathbf{q}) \leq \omega^{i-1} f(\mathbf{q}), \forall \mathbf{q} \in \mathcal{Q}$. For each $\mathbf{q} \in \mathcal{Q}$, we then have
\begin{align} \nonumber
\omega^{i+1} f(\mathbf{q}) &= \omega \, \omega^i  f(\mathbf{q}) \\ \nonumber
&= \min\{q^{0}, \, c \, (1 - q^{0}) + \min_{\pmb{\varphi}} K_{\omega^i  f}(\pmb{\varphi}, \mathbf{q})\} \\ \nonumber
&\leq \min\{q^{0}, \, c \, (1 - q^{0}) + \min_{\pmb{\varphi}} K_{\omega^{i-1} f}(\pmb{\varphi}, \mathbf{q})\} \\ \nonumber
&= \omega \, \omega^{i-1}  f(\mathbf{q}) = \omega^i f(\mathbf{q}).
\end{align}

Since $\omega^{i+1} f(\mathbf{q}) \leq \omega^i f(\mathbf{q}), \forall \mathbf{q} \in \mathcal{Q}, \forall i \geq 0$ and by definition, $\omega^i f(\mathbf{q}) \geq 0, \forall \mathbf{q} \in \mathcal{Q}, \forall i \geq 0$, $\omega^i f(\mathbf{q}), \mathbf{q} \in \mathcal{Q}$ converges to a function in $\mathcal{S}$ as $i \rightarrow \infty$. Since by Proposition 3-(a), $J(\mathbf{q})$ is the unique fixed point of the mapping $\omega$, we have $\omega^i f(\mathbf{q}) \rightarrow J(\mathbf{q})$ as $i \rightarrow \infty$. This leads to a numerical method to compute $J(\mathbf{q}), \mathbf{q} \in \mathcal{Q}$ \cite{Veeravalli01,Veeravalli93}. In this method, firstly a representative set of points in the simplex $\mathcal{Q}$ is chosen. Let this set of points be denoted with $\mathcal{G}$, i.e., $\mathcal{G} \subset \mathcal{Q}$. Then, considering $i$ as the iteration number, at each iteration, $i$ increased by one and $\omega^i f(\mathbf{q}), \mathbf{q} \in \mathcal{G}$ is obtained from $\omega^{i-1} f(\mathbf{q}), \mathbf{q} \in \mathcal{G}$ using the mapping $\omega$. When the difference in values between two consecutive iterations, i.e., $|\omega^i f(\mathbf{q}) - \omega^{i-1} f(\mathbf{q})|$ for each $\mathbf{q} \in \mathcal{G}$, is below a certain threshold, say $\epsilon$, then the iterations can be stopped and after the last iteration, $J(\mathbf{q}), \mathbf{q} \in \mathcal{G}$ is approximated as $\omega^i f(\mathbf{q}), \mathbf{q} \in \mathcal{G}$. After then, for each $\mathbf{q} \in \mathcal{G}$, the optimal thresholds can be approximated as follows:
\begin{gather} \label{eq:opt_threshold_v1}
\pmb{\varphi}_{\text{opt}}(\mathbf{q}) = \arg \min_{\pmb{\varphi}} K_J(\pmb{\varphi}, \mathbf{q}).
\end{gather}
Then, from \eqref{eq:A_q} and \eqref{eq:K_B}, we have $A(\mathbf{q}) = \min_{\pmb{\phi} \, \subset \, \pmb{\Phi}} K_J(\pmb{\phi}, \mathbf{q}) = K_J(\pmb{\varphi}_{\text{opt}}(\mathbf{q}),\mathbf{q}), \forall \mathbf{q} \in \mathcal{G}$. The optimal stopping time can then be approximately determined using \eqref{eq:inf_horizon_sol_v0} where the sufficient statistics at each time can be updated as follows:
\begin{gather} \label{eq:recursion_v3}
\mathbf{q}_{k} = \frac{\mathbf{g}_{k}(\mathbf{u}_k, \pmb{\varphi}_{\text{opt}}(\mathbf{q}_{k-1}), \mathbf{q}_{k-1})}{h_k(\mathbf{u}_k, \pmb{\varphi}_{\text{opt}}(\mathbf{q}_{k-1}), \mathbf{q}_{k-1})}.
\end{gather}
Note that \eqref{eq:recursion_v3} is obtained using \eqref{eq:recursion_v7} and the initial sufficient statistics are $q_0^0 = 1$ and $q_0^n = 0, \forall n \in \{1, 2, \dots, L\}$.

Hence, the implementation of the optimal solution consists of two stages. In the first stage, $J(\mathbf{q})$, $\pmb{\varphi}_{\text{opt}}(\mathbf{q})$, and $A(\mathbf{q})$ are approximately calculated for all $\mathbf{q} \in \mathcal{G}$ as described above and summarized in Algorithm \ref{alg:offline}. This stage needs to be performed offline before taking any measurements. Note that the computational complexity of Algorithm \ref{alg:offline} increases as the set $\mathcal{G}$ is enlarged and the threshold $\epsilon$ is decreased. Hence, for a better approximation of the optimal solution, more computational resources need to be employed in the offline stage. After the offline stage, measurements are obtained sequentially and the optimal procedures for change detection are employed at sensors and the fusion center. This corresponds to an online stage in which both the fusion center and sensors need to update sufficient statistics at each time. Hence, having a simpler update formula is useful to simplify the implementation. Towards this goal, we use the following transformation for all ${n \in \{0, 1, \dots, L\}}$:
\begin{gather} \label{eq:transf_v1}
p_{k}^{n} = \frac{q_{k}^{n}}{\rho \, q_{k}^{0}} \, \Longleftrightarrow \, q_{k}^{n} = \frac{p_{k}^{n}}{\sum_{m=0}^{L} p_{k}^{m}}
\end{gather}
and obtain ${\mathbf{p}_k \triangleq [p_{k}^{0}, p_{k}^{1} \dots, p_{k}^{L}]}$ as the new sufficient statistics.

\begin{algorithm}[t]\small
\caption{\small Implementation of the optimal solution: Offline stage}
\label{alg:offline}
\baselineskip=0.5cm
\begin{algorithmic}[1]
\STATE Initialization: $i \gets 0$, $d \gets \infty$, choose an $\epsilon > 0$, choose a set $\mathcal{G}$ of representative points in $\mathcal{Q}$.
\WHILE {$d > \epsilon$}
    \STATE $i \gets i+1$.
    \STATE $\omega^i f(\mathbf{q}) \gets \omega \, \omega^{i-1} f(\mathbf{q})$, $\mathbf{q} \, \in \, \mathcal{G}$.
    \STATE $d \gets \max_{\mathbf{q} \, \in \, \mathcal{G}} \{ \, |\omega^i f(\mathbf{q}) - \omega^{i-1} f(\mathbf{q})| \, \}$.
\ENDWHILE
\STATE $J(\mathbf{q}) \gets \omega^i f(\mathbf{q}), \forall \mathbf{q} \in \mathcal{G}$.
\IF {the centralized model is used}
    \STATE $A(\mathbf{q}) \gets K_J(\pmb{\phi}, \mathbf{q})$ where $\pmb{\phi}$ were initially chosen as the identity functions.
\ELSIF{the decentralized model is used}
    \STATE $\pmb{\varphi}_{\text{opt}}(\mathbf{q}) \gets \arg \min_{\pmb{\varphi}} K_J(\pmb{\varphi}, \mathbf{q}), \forall \mathbf{q} \in \mathcal{G}$.
    \STATE $A(\mathbf{q}) \gets K_J(\pmb{\varphi}_{\text{opt}}(\mathbf{q}), \mathbf{q}), \forall \mathbf{q} \in \mathcal{G}$.
\ENDIF
\end{algorithmic}
\end{algorithm}

From \eqref{eq:transf_v1}, we have $p_{k}^{0} = 1/\rho, \forall k \geq 1$. Then, $q_{k}^{n}$ can be expressed as follows:
\begin{gather} \label{eq:qkn}
q_{k}^{n} = \frac{\rho \, p_{k}^{n}}{1 + \rho \, \sum_{m=1}^{L} p_{k}^{m}}.
\end{gather}
The optimal thresholds obtained in \eqref{eq:opt_threshold_v1} can then be written as functions of the new sufficient statistics as $\hat{\pmb{\varphi}}_{\text{opt}} \triangleq [\hat{\pmb{\varphi}}_{\text{opt},1}, \dots, \hat{\pmb{\varphi}}_{\text{opt},L}]$ where ${\hat{\pmb{\varphi}}_{\text{opt},\ell} \triangleq [\hat{\varphi}_{\ell,0}, \hat{\varphi}_{\ell,1}, \dots, \hat{\varphi}_{\ell,U_\ell}]}$ and calculated based on \eqref{eq:qkn} as follows:
\begin{gather} \label{eq:opt_threshold_v2}
\hat{\pmb{\varphi}}_{\text{opt}}(\mathbf{p}) = {\pmb{\varphi}}_{\text{opt}}\bigg(\frac{\rho \, \mathbf{p}}{1 + \rho \, \sum_{m=1}^{L} p^{m}}\bigg).
\end{gather}
Then, the set of optimal thresholds at each time $k$ can also be obtained as ${\pmb{\varphi}}_k = \hat{\pmb{\varphi}}_{\text{opt}}(\mathbf{p}_{k-1})$.

Using \eqref{eq:recursion_v3}, \eqref{eq:transf_v1}, and \eqref{eq:g_inf}, the recursion for $p_{k}^{n}, n \in \{1,2,\dots,L\}$ is obtained as
\begin{align} \nonumber
p_{k}^{n}
&= \frac{q_{k}^{n}}{\rho \, q_{k}^{0}} = \frac{g_{k}^{n}(\mathbf{u}_k, \pmb{\varphi}_{\text{opt}}(\mathbf{q}_{k-1}), \mathbf{q}_{k-1})}{\rho \, g_{k}^{0}(\mathbf{u}_k, \pmb{\varphi}_{\text{opt}}(\mathbf{q}_{k-1}), \mathbf{q}_{k-1})} \\ \nonumber
&= \delta_{k}^{n} \, \frac{1-\rho_{\pi_{n},\pi_{n+1}}}{1-\rho} \, \frac{\sum_{m=0}^{n} q_{k-1}^{m} \, e_{m}^{n}}{\rho \, q_{k-1}^{0}}
\\ \label{eq:recursion_pkn}
&= \delta_{k}^{n} \, \frac{1-\rho_{\pi_{n},\pi_{n+1}}}{1-\rho} \, \left(\sum_{m=0}^{n} p_{k-1}^{m} \, e_{m}^{n}\right),
\end{align}
where
\begin{gather} \nonumber
\delta_{k}^{n} \triangleq \bigg( \sum_{\Pi} \, \prod_{i=1}^{n} \mathbb{P}_1\big({\varphi}_{k,\pi_{i}, u_{k,\pi_{i}}} < L(z_{\pi_{i}}) \leq {\varphi}_{k,\pi_{i},(u_{k,\pi_{i}}+1)} \big)
 \prod_{i=n+1}^{L} \mathbb{P}_0\big({\varphi}_{k,\pi_{i}, u_{k,\pi_{i}}} < L(z_{\pi_{i}}) \leq {\varphi}_{k,\pi_{i},(u_{k,\pi_{i}}+1)}\big) \bigg) \\ \label{eq:delta}
\times \left(\sum_{\Pi} \, \prod_{i=1}^{L} \mathbb{P}_0\big({\varphi}_{k,\pi_{i}, u_{k,\pi_{i}}} < L(z_{\pi_{i}}) \leq {\varphi}_{k,\pi_{i},(u_{k,\pi_{i}}+1)}\big) \right)^{-1}.
\end{gather}
Note that the initial values of the sufficient statistics are $p_{0}^{0} = {1}/{\rho}$ and $p_{0}^{n} = 0, \forall n \in \{1, 2, \dots, L\}$.

We observe that compared to \eqref{eq:recursion_v3}, we have a relatively simpler recursive formula in \eqref{eq:recursion_pkn} and since $p_{k}^{0}$ is constant over time, at each time, update of an $L$-dimensional vector is needed instead of $L+1$. Further, through \eqref{eq:recursion_pkn} and \eqref{eq:delta}, we observe that the update of the sufficient statistics requires the knowledge of all sensor messages. Since the update must be done at both sensors and the fusion center, feedback from the fusion center is needed to implement the optimal solution in the decentralized setting.

Furthermore, using \eqref{eq:qkn}, we have
\begin{gather} \label{eq:temp}
q_{k}^{0} = \frac{1}{1 + \rho \, \sum_{m=1}^{L} p_{k}^{m}}.
\end{gather}
Then, based on \eqref{eq:inf_horizon_sol_v0} and \eqref{eq:temp}, the optimal stopping rule can be rewritten as follows:
\begin{gather} \label{eq:inf_horizon_sol_v14}
\tau = \min\bigg\{k \in \mathbb{N}: \sum_{m=1}^{L} p_{k}^{m} \geq \frac{1-\hat{A}(\mathbf{p}_k)}{\rho \, (c + \hat{A}(\mathbf{p}_k))}\bigg\},
\end{gather}
where
\begin{gather}\nonumber
\hat{A}(\mathbf{p}_k) \triangleq A\bigg(\frac{\rho \, \mathbf{p}_k}{1 + \rho \, \sum_{m=1}^{L} p_k^{m}}\bigg).
\end{gather}

Finally, we note that in the centralized setting, the local message functions are replaced with the identity functions. Then, we have $u_{k,\ell} = z_{k,\ell}, \forall \ell \in \{1, 2, \dots, L\}$ and $\forall k \geq 1$. After this change, all the previously stated results (except on the structure of the optimal local message functions) hold true and the corresponding test structure is obtained as in \eqref{eq:inf_horizon_sol_v14}. Further, the recursion of sufficient statistics is obtained as in \eqref{eq:recursion_pkn} after replacing $\delta_{k}^{n}$ with
\begin{gather} \label{eq:delta_cent}
\delta_{k}^{n} = \bigg( \sum_{\Pi} \, \prod_{i=1}^{n} f_1(z_{k,\pi_{i}}) \prod_{i=n+1}^{L} f_0(z_{k,\pi_{i}}) \bigg)
 \bigg(\sum_{\Pi} \, \prod_{i=1}^{L} f_0(z_{k,\pi_{i}}) \bigg)^{-1}.
\end{gather}

Based on the discussion we have above on the online implementation stage of the optimal solution, we summarize the procedures at a sensor and the fusion center in Algorithm \ref{alg:online_sensor} and Algorithm \ref{alg:online_fc}, respectively in both centralized and decentralized settings. Note that at a sensor, the update of sufficient statistics is performed with one-time-instant delay (see lines 7-10 in Algorithm \ref{alg:online_sensor}) compared to the fusion center. This is due to the fact that a sensor has access to messages of other sensors via feedback from the fusion center with one-time-instant delay.

\begin{algorithm}[t]\small
\caption{\small Implementation of the optimal solution: Online stage at sensor $\ell$}
\label{alg:online_sensor}
\baselineskip=0.5cm
\begin{algorithmic}[1]
\STATE Initialization: $k \gets 0$, $p_{0}^{0} \gets {1}/{\rho}$, $p_{0}^{n} \gets 0, \forall n \in \{1, 2, \dots, L\}$.
\WHILE {$k < \tau$}
    \STATE $k \gets k+1$.
    \IF {the centralized model is used}
        \STATE Send $z_{k,\ell}$ to the fusion center.
    \ELSIF{the decentralized model is used}
        \IF {$k > 1$}
            \STATE Compute $\delta_{k-1}^n$ using \eqref{eq:delta}, $\forall n \in \{1,2,\dots,L\}$.
            \STATE Compute $p_{k-1}^n$ using \eqref{eq:recursion_pkn}, $\forall n \in \{1,2,\dots,L\}$.
        \ENDIF
        \STATE ${\pmb{\varphi}}_k \gets {\pmb{\varphi}}_{\text{opt}}\big(\frac{\rho \, \mathbf{p}_{k-1}}{1 + \rho \, \sum_{m=1}^{L} p_{k-1}^{m}}\big)$.
        \IF {${\varphi}_{k,\ell,i} < L(z_{k,\ell}) \leq {\varphi}_{k,\ell,i+1}$}
            \STATE Send $u_{k,\ell} = i$ to the fusion center using $\lceil \log_2(U_\ell) \rceil$ bits.
        \ENDIF
    \ENDIF
\ENDWHILE
\end{algorithmic}
\end{algorithm}

\begin{algorithm}[t]\small
\caption{\small Implementation of the optimal solution: Online stage at the fusion center}
\label{alg:online_fc}
\baselineskip=0.5cm
\begin{algorithmic}[1]
\STATE Initialization: $k \gets 0$, $p_{0}^{0} \gets {1}/{\rho}$, $p_{0}^{n} \gets 0, \forall n \in \{1, 2, \dots, L\}$.
\WHILE {$k < \tau$}
    \STATE $k \gets k+1$.
    \IF {the centralized model is used}
        \STATE Compute $\delta_k^n$ using \eqref{eq:delta_cent}, $\forall n \in \{1,2,\dots,L\}$.
    \ELSIF{the decentralized model is used}
        \STATE ${\pmb{\varphi}}_{k} \gets {\pmb{\varphi}}_{\text{opt}}\big(\frac{\rho \, \mathbf{p}_{k-1}}{1 + \rho \, \sum_{m=1}^{L} p_{k-1}^{m}}\big)$.
        \STATE Compute $\delta_k^n$ using \eqref{eq:delta}, $\forall n \in \{1,2,\dots,L\}$.
    \ENDIF
    \STATE Compute $p_k^n$ using \eqref{eq:recursion_pkn}, $\forall n \in \{1,2,\dots,L\}$.
    \STATE $\hat{A}(\mathbf{p}_k) \gets A\bigg(\frac{\rho \, \mathbf{p}_k}{1 + \rho \, \sum_{m=1}^{L} p_k^{m}}\bigg)$.
    \IF {$\sum_{m=1}^{L} p_{k}^{m} \geq \frac{1-\hat{A}(\mathbf{p}_k)}{\rho \, (c + \hat{A}(\mathbf{p}_k))}$}
        \STATE $\tau \gets k$.
    \ENDIF
\ENDWHILE
\STATE Declare the change and stop the procedure.
\end{algorithmic}
\end{algorithm}

\section{The Rare Change Regime} \label{sec:rare_change}

The optimal stopping rule is given in \eqref{eq:inf_horizon_sol_v14}. However, further analytical investigation of the optimal solution is difficult. Although some numerical techniques can be used to approximately compute the infinite-horizon cost-to-go function $J(\mathbf{q})$ and the optimal thresholds of the local monotone LR quantizers as summarized in Algorithm \ref{alg:offline} and to implement the online optimal change detection procedures as summarized in Algorithms \ref{alg:online_sensor}-\ref{alg:online_fc}, such techniques may not be preferable in a practical setting since (i) the offline stage described in Algorithm \ref{alg:offline} is computationally challenging especially when the goal is to precisely obtain the optimal solution (see lines 3-7 \& 12 of Algorithm \ref{alg:offline} and consider a small $\epsilon$ and a large set $\mathcal{G}$), (ii) in the decentralized setting, the online stage requires update of sufficient statistics at each time at all sensors (see lines 7-11 in Algorithm \ref{alg:online_sensor}) in addition to the fusion center. Moreover, the optimal solution becomes practically infeasible in the decentralized setting (cf. Remark 2). In this section, to eliminate such difficulties, we study a practically important special regime where the change occurs rarely, i.e., $\rho \rightarrow 0$ \cite{Poor08,Basseville93,Tartakovsky08,Veeravalli01,Premkumar09}. In this regime, we propose and analyze some simpler solutions.

Firstly, from \eqref{eq:opt_threshold_v2}, as $\rho \rightarrow 0$, we have
\begin{gather}  \nonumber
\hat{\pmb{\varphi}}_{\text{opt}}(\mathbf{p}) = {\pmb{\varphi}}_{\text{opt}}(\mathbf{0})
\end{gather}
for all $\mathbf{p}$, which shows that the optimal thresholds of the local message functions in the rare change regime are independent of the sufficient statistics, equivalently past sensor messages. We have previously stated that for each sensor, the optimal local message function is a monotone LR quantizer with stationary threshold functions of the sufficient statistics. These two results together indicate that monotone LR quantizers with constant thresholds are optimal in the rare change regime. However, at this point, we still do not have an efficient method to determine these thresholds. In the following, we firstly derive the structure of $\tau$ given in \eqref{eq:inf_horizon_sol_v14} as $\rho \rightarrow 0$. We then present a multichart scheme and show an asymptotic optimality of the corresponding test as its false alarm rate goes to zero. Then, how to optimally choose the quantization thresholds in this asymptotic regime will be clear.

The following proposition presents the limiting form of the optimal test structure in the rare change regime and a way to determine threshold of the corresponding test to set an upper bound for the false alarm probability.

\textbf{Proposition 4:} (a) As $\rho \rightarrow 0$, $\tau$ given in \eqref{eq:inf_horizon_sol_v14} converges to $\nu(\beta)$, given by
\begin{gather} \label{eq:rare_change}
\nu(\beta) = \min \bigg\{k \in \mathbb{N}: \log{\sum_{m=1}^{L} p_{k}^{m}} \geq \beta \bigg\},
\end{gather}
where $\beta$ is a properly chosen test threshold.

(b) If $\beta = \log(\frac{1}{\rho \, \alpha})$ where $\alpha \in (0, 1)$, then PFA$\left(\nu(\beta)\right) \leq \alpha$.

\begin{proof}
(a) See Appendix \ref{sec:proof_rare}.

(b) Using \eqref{eq:temp}, \eqref{eq:rare_change} can be rewritten as
\begin{gather} \label{eq:rare_change_v3} \nonumber
\nu(\beta) = \inf \bigg\{k \in \mathbb{N}: q_{k}^{0} \leq \frac{1}{1 + \rho \, \exp{(\beta)}} \bigg\}.
\end{gather}
Then, the false alarm probability of $\nu(\beta)$ can be written as follows:
\begin{gather} \label{eq:pfa} \nonumber
\text{PFA}(\nu(\beta)) = \mathbb{P}(S_{\nu(\beta)} = 0)
 = \mathbb{E}\big[q_{\nu(\beta)}^{0}\big] \leq \frac{1}{1 + \rho \, \exp{(\beta)}}.
\end{gather}
If $\beta = \log({1}/{(\rho \, \alpha)})$, then $\text{PFA}\left(\nu(\beta)\right) \leq \alpha$.
\end{proof}

Recall that we have initially assumed all change patterns are equally likely (cf. \eqref{eq:uni_perm}) and arrived at $\nu(\beta)$ as the limiting form of the optimal stopping rule as $\rho \rightarrow 0$. While we find it difficult to show the optimality or asymptotic optimality of $\nu(\beta)$ where $\beta$ is chosen as in Proposition 4-(b), after removing the uniform-prior assumption given in \eqref{eq:uni_perm} and proposing a multichart scheme as explained in the next subsection, we show an order-1 asymptotic optimality of the corresponding multichart test.

Note that although \eqref{eq:rare_change} with $\beta = \log({1}/{(\rho \, \alpha)})$ is not necessarily optimal, since (i) it has a simple thresholding procedure with an easily determined threshold at the fusion center, (ii) it corresponds to a practically feasible scheme with no feedback requirement for the past sensor messages unlike the optimal solution where sensors need feedback to update the sufficient statistics, we consider \eqref{eq:rare_change} as our proposed test for the uniform-prior scheme. After we study the multichart scheme in the next subsection and insights we will gain from this discussion, we will propose a reasonable suboptimal method to determine the quantization thresholds at sensors (cf. Remark 3).

\subsection{Multichart Scheme} \label{sec:multichart}

We now assume that the change propagation pattern is an unknown non-random quantity and propose that the decision statistic at each time is determined by the change propagation pattern that maximizes it. In particular, it can be checked that assuming (wrongly or correctly) a specific change propagation pattern is true, say $\Pi = \Pi_j$, and following the same solution methodology we have in this paper, all the results presented in the previous sections hold true after replacing $g_{k}^{n}, n \in \{0,1,\dots,L\}$ in \eqref{eq:gkn_tmp} with $g_{k,j}^n, n \in \{0,1,\dots,L\}$ and all the other related terms correspondingly to indicate that $\Pi = \Pi_j$ is assumed.
In this case, $\delta_{k}^{n}$ in \eqref{eq:delta} is reduced to
\begin{gather} \label{eq:delta_reduced}
\delta_{k,j}^{n} \triangleq  \prod_{i=1}^{n} \frac{\mathbb{P}_1\big({\varphi}_{k,\pi_{i}, u_{k,\pi_{i}}} < L(z_{\pi_{i}}) \leq {\varphi}_{k,\pi_{i},(u_{k,\pi_{i}}+1)}\big)}{\mathbb{P}_0\big({\varphi}_{k,\pi_{i}, u_{k,\pi_{i}}} < L(z_{\pi_{i}}) \leq {\varphi}_{k,\pi_{i},(u_{k,\pi_{i}}+1)}\big)},
\end{gather}
and similar to \eqref{eq:recursion_pkn}, we end up with the following recursion of sufficient statistics:
\begin{gather} \label{eq:recursion_pkn_reduced}
p_{k,j}^{n} = \delta_{k,j}^{n} \, \frac{1-\rho_{\pi_{n},\pi_{n+1}}}{1-\rho} \, \bigg(\sum_{m=0}^{n} p_{k-1,j}^{m} \, e_{m}^{n}\bigg).
\end{gather}
Then, the corresponding test in the rare change regime is obtained as follows:
\begin{gather} \label{eq:tau_j}
\nu_j(\beta) \triangleq \min \bigg\{k \in \mathbb{N}: \log{\sum_{m=1}^{L} p_{k,j}^{m}} \geq \beta \bigg\}.
\end{gather}

We consider to maximize the decision statistic at any time over all possible change propagation patterns and hence propose the following test:
\begin{align} \nonumber
\nu_{\min}(\beta) &= \min \bigg\{k \in \mathbb{N}:  \max_{\Pi_j} \, \log{\sum_{m=1}^{L} p_{k,j}^{m}} \geq \beta \bigg\} \\ \label{eq:multi_chart_v1}
&= \min \, \{ \nu_j(\beta), ~ j = 1, 2, \dots, L! \}.
\end{align}
Based on \eqref{eq:multi_chart_v1}, we observe that the test in \eqref{eq:tau_j} is performed for all possible patterns simultaneously and the change is declared  at the first time any test crosses a predetermined threshold. We thus call \eqref{eq:multi_chart_v1} a multichart test.

\subsection{Asymptotic Optimality}

Next, we provide a first-order asymptotic optimality result for \eqref{eq:multi_chart_v1} for small values of false alarm probability.  Before we proceed our discussion in the decentralized setting, we find it useful firstly to analyze the centralized counterpart of $\nu_{\min}(\beta)$ where the fusion center has access to the exact measurements instead of outputs of local quantizers. Let the corresponding test in the centralized setting be denoted with $\nu_{\min}^c(\beta)$, given by
\begin{gather} \label{eq:multi_chart_cent}
\nu_{\min}^c(\beta) = \min \{ \nu_j^c(\beta), ~ j = 1, 2, \dots, L! \},
\end{gather}
where $\nu_{j}^{c}(\beta)$ denotes the centralized counterpart of $\nu_j(\beta)$. Note that in the centralized setting,
\eqref{eq:recursion_pkn_reduced} gives the recursive formula for $p_{k,j}^{n}$ where $\delta_{k,j}^{n}$ is calculated as follows:
\begin{gather} \label{eq:delta_reduced_cent}
\delta_{k,j}^{n} =  \prod_{i=1}^{n} \frac{f_1(z_{k,\pi_{i}})}{f_0(z_{k,\pi_{i}})}.
\end{gather}

We observe that if $\Pi$ is known, i.e., if $\Pi = \Pi_j$ is true, then $\nu_{j}^{c}$ has the identical form with the test proposed in \cite{Raghavan10} where an order-1 asymptotic optimality result is shown. Using this result, we show in the following proposition that $\nu_{\min}^c$ is asymptotically optimal as $\alpha \rightarrow 0$ among all tests with false alarm probability upper bounded by $\alpha$ under some sufficient conditions on $D(f_1, f_0)$ where
\begin{gather} \nonumber
D(f_1, f_0) = \int f_1(x) \, \log\bigg(\frac{f_1(x)}{f_0(x)}\bigg) \, dx
\end{gather}
denotes the K-L information distance between the post- and pre-change densities.

\textbf{Proposition 5:} Let $\Lambda_\alpha \triangleq \{\tau: \text{PFA}(\tau) \leq \alpha \}$ be the class of stopping rules with false alarm probability upper bounded by $\alpha$ and let $\beta = \log(\frac{1}{\rho \, \alpha})$. If
\begin{gather} \label{eq:cond_opt}
\log\big(1-\rho + (L-1) (1 - \lambda)\big) < D(f_1, f_0) < \infty,
\end{gather}
then $\nu_{\min}^{c}(\beta)$ is order-1 asymptotically optimal as $\alpha \rightarrow 0$ for any $\rho \in (0, 1]$. That is,
\begin{gather} \label{eq:asym_opt}
\lim_{\alpha \, \rightarrow \, 0} \frac{\inf_{\tau \, \in \, \Lambda_\alpha} \text{ADD}(\tau)}{\mathbb{E}\,[\nu_{\min}^{c}(\beta)]} = 1,
\end{gather}
where
\begin{gather} \label{eq:adQ_lower_bnd}
\inf_{\tau \, \in \, \Lambda_\alpha} \text{ADD}(\tau) \geq \frac{|\log(\alpha)| (1 + o(1))}{L \, D(f_1, f_0) + |\log(1 - \rho)|} ~~ \text{as} ~ \alpha \rightarrow 0,
\end{gather}
and
\begin{gather} \label{eq:add_upper_bnd}
\mathbb{E}\,[\nu_{\min}^{c}(\beta)] \leq \frac{|\log(\rho \, \alpha)| (1 + o(1))}{L \, D(f_1, f_0) + |\log(1 - \rho)|} ~~ \text{as} ~ \alpha \rightarrow 0.
\end{gather}

\begin{proof}
Let $\beta = \log(\frac{1}{\rho \, \alpha})$ as stated in the proposition and let
\begin{gather} \nonumber
\tilde{\Lambda} \triangleq \big\{\nu_{j}^{c}(\beta), ~ j = 1, 2, \dots, L! \big\}.
\end{gather}
By definition (cf. \eqref{eq:multi_chart_cent}), $\nu_{\min}^{c}(\beta) \in \tilde{\Lambda}$. Following the same arguments given in the proof of Proposition 4-(b), it can be shown that $\tilde{\Lambda} \subset \Lambda_\alpha$. Then by \cite[Proposition 4]{Raghavan10}, for any stopping rule in $ \tau \in \Lambda_\alpha$, a lower bound on the ADD as $\alpha \rightarrow 0$ is obtained as in \eqref{eq:adQ_lower_bnd}. Let $\Pi_t$ denote the true change propagation pattern, i.e., $\Pi = \Pi_t$, and $\nu_{t}^{c}(\beta)$ be the corresponding stopping time. By \cite[Theorem 3]{Raghavan10}, for $\nu_{t}^{c}(\beta)$, an upper bound on the ADD as $\alpha \rightarrow 0$ is obtained as the RHS of the inequality in \eqref{eq:add_upper_bnd} if the condition given in \eqref{eq:cond_opt} is satisfied\footnote{\eqref{eq:cond_opt} directly comes from the condition given in \cite[Theorem 3]{Raghavan10} when the inter-sensor change parameters are the same, i.e., $\lambda_{\ell, \ell -1} = \lambda, \forall \ell = 2, \cdots, L$. Further, if $\lambda \rightarrow 1$, \eqref{eq:cond_opt} reduces to a mild condition: $D(f_1, f_0)$ must be positive and finite.}. Due to \eqref{eq:multi_chart_cent}, $\nu_{\min}^{c}(\beta) \leq \nu_{t}^{c}(\beta)$ always holds, thus \eqref{eq:add_upper_bnd} follows. Based on \eqref{eq:adQ_lower_bnd} and \eqref{eq:add_upper_bnd}, we obtain \eqref{eq:asym_opt}, i.e., $\nu_{\min}^{c}(\beta)$ is the optimal test among all tests whose false alarm probability is upper bounded by $\alpha$ in the asymptotic regime where $\alpha \rightarrow 0$.
\end{proof}

We have previously argued that the optimal local message functions are monotone LR quantizers with constant thresholds. Due to the quantizer at each sensor $\ell$, a range of LRs is mapped to each message in the alphabet $\{0, 1, \dots, U_\ell -1\}$ under each density. The corresponding pmfs on sensor messages are denoted with $P_{0,\ell}^{u}$ and $P_{1,\ell}^{u}$  under pre- and post-change densities, respectively and calculated as follows:
\begin{gather} \label{eq:induced_pmf0}
P_{0,\ell}^{u}(i) = \mathbb{P}_0({\varphi}_{\ell,i} < L(z_\ell) \leq {\varphi}_{\ell,i+1}),\\ \label{eq:induced_pmf1}
P_{1,\ell}^{u}(i) = \mathbb{P}_1({\varphi}_{\ell,i} < L(z_\ell) \leq {\varphi}_{\ell,i+1}),
\end{gather}
$\forall i \in \{0, 1, \dots, U_{\ell-1}\}$, where $z_\ell$ denotes a generic observation and $\{{\varphi}_{\ell,i}, i = 0, 1, \dots, U_{\ell-1}\}$ denote the constant thresholds at sensor $\ell$. Let
\begin{gather} \nonumber
D(P_{1,\ell}^{u}, P_{0,\ell}^{u}) = \sum_{i=0}^{U_\ell-1} P_{1,\ell}^{u}(i) \, \log\bigg(\frac{P_{1,\ell}^{u}(i)}{P_{0,\ell}^{u}(i)}\bigg)
\end{gather}
be the K-L information distance between the pmfs $P_{1,\ell}^{u}$ and $P_{0,\ell}^{u}$.

Proposition 5 shows that for small values of PFA, the ADD of the multichart test in the centralized setting is inversely proportional to $L \, D(f_1, f_0)$ under certain conditions. With constant thresholds at sensors, we have a setup equivalent to a centralized setup in which $u_{k,1}, u_{k,2}, \dots, u_{k,L}$ are the i.i.d. ``measurements" at time $k$ \cite{Veeravalli01}. Then, as an application of Proposition 5, as $\alpha \, \rightarrow \, 0$, the ADD of the decentralized multichart test is inversely proportional to $\sum_{\ell=1}^{L} D(P_{1,\ell}^{u}, P_{0,\ell}^{u})$ under certain conditions. Hence, in order to minimize the ADD in the asymptotic regime, $D(P_{1,\ell}^{u}, P_{0,\ell}^{u})$ should be maximized at each sensor $\ell$. Then, choosing the quantization thresholds of the monotone LR quantizer at sensor $\ell$ so as to maximize $D(P_{1,\ell}^{u}, P_{0,\ell}^{u})$ is asymptotically optimal. This presents a simple way to choose the quantization thresholds. A similar result is obtained in \cite{Tartakovsky08} where a decentralized change detection problem is studied assuming the same change points for all sensors.

Due to symmetry, we assume the same alphabet size for all sensors, i.e., ${U_1 = U_2 = \dots = U_L}$. Then because the pre- and post-change pdfs $f_0$ and $f_1$ are common across sensors, the asymptotically optimal thresholds are identical at all sensors. Let $\Lambda_\alpha \triangleq \{\Omega = (\pmb{\phi},\tau): \text{PFA}(\tau) \leq \alpha\}$ be the class of policies consisting of monotone LR quantizers with constant thresholds, denoted with $\pmb{\phi}$, and the stopping rules with probability of false alarm not bigger than $\alpha$. Moreover, let ${\pmb{\varphi}}^*$ denote the set of asymptotically optimal thresholds and let the corresponding (common) pmfs be denoted with $P_{1}^{u^*}$ and $P_{0}^{u^*}$. Then, we have the following corollary to Proposition 5. 

\textbf{Corollary 1:} Let $\beta = \log(\frac{1}{\rho \, \alpha})$. If the local quantization thresholds are chosen as ${\pmb{\varphi}}^*$ and
\begin{gather} \label{eq:cond_opt_dc}
\log\big(1-\rho + (L-1) (1 - \lambda)\big) < D(P_{1}^{u^*}, P_{0}^{u^*}) < \infty,
\end{gather}
then
\begin{gather} \label{eq:decent_perf}
{\inf_{\Omega \, \in \, \Lambda_\alpha} \text{ADD}(\tau)} \sim {\mathbb{E}\,[\nu_{\min}(\beta)]}
\sim \frac{|\log(\alpha)|}{L \, D(P_{1}^{u^*}, P_{0}^{u^*}) + |\log(1 - \rho)|},
\end{gather}
where $a(\alpha) \sim b(\alpha)$ denotes $\lim_{\alpha \, \rightarrow \, 0} \frac{a(\alpha)}{b(\alpha)} = 1$.

{\textit{{Remark 3:}} With the motivation of asymptotically optimal local quantizers obtained for the multichart scheme and since the numerical results in Section \ref{sec:sim} indicate that the average detection delay is inversely proportional to the K-L information distance, we propose that the quantization thresholds in the decentralized uniform-prior scheme are also selected to maximize the K-L information distance between the post- and pre-change pmfs. Although this choice is suboptimal compared to the optimal procedure explained in Section \ref{sec:inf_horizon}, it reduces the computational complexity of determining the quantization thresholds, so considerably simplifies the implementation in practice.

Based on the discussion above, we summarize the procedure at a sensor node for both the uniform-prior and multichart schemes in both the centralized setting and the decentralized setting with the uniform-in-time sampling scheme in Algorithm \ref{alg:sensor} (see lines 4-10). Note that the local thresholds are initially determined using the described method and there is no need to compute and update sufficient statistics at a sensor. Further, we summarize the procedure at the fusion center in both the centralized setting and decentralized setting with the uniform-in-time sampling scheme in Algorithm \ref{alg:uniprior} (see lines 4-8 \& 17-20) for the uniform-prior scheme and in Algorithm \ref{alg:multichart} (see lines 10-16 \& 21-25) for the multichart scheme. Computationally, the multichart test requires to perform $L!$ tests simultaneously where each test has a simple recursive formula for sufficient statistics. On the other hand, the uniform-prior test requires to perform a single test while the update of sufficient statistics requires summations over $L!$ possible patterns. Hence, both tests are computationally demanding especially for large networks consisting of many sensors.

\section{Decentralized Implementation Using Level-Crossing Sampling} \label{sec:event_based}

Since $D(f_{1}, f_{0}) > D(P_{1}^{u^*}, P_{0}^{u^*})$ is always true, there is a clear performance gap between the proposed centralized and decentralized tests (cf. \eqref{eq:add_upper_bnd} and \eqref{eq:decent_perf}). In the decentralized setting, we have assumed until this point that sensors use the conventional uniform-in-time sampling and a finite-size alphabet for sampling and quantization of LRs. To make efficient use of the communication resources, we now propose that sensors use an event-triggered sampling scheme called level-crossing sampling with hysteresis (LCSH) \cite{Yilmaz15,Yilmaz18,Necip18}.

With uniform-in-time sampling, each sensor communicates with the fusion center at each discrete time. It is highly probable that many non-informative transmissions occur, hence energy and bandwidth resources are not effectively used. Especially in a case where change happens rarely, since the pre-change duration is expected to be much longer compared to the post-change duration, periodic transmission may lead to significant waste of resources. On the contrary, in event-triggered sampling schemes, a sensor communicates with the fusion center only when a well-defined event occurs indicating a significant change in local statistics. Hence, such schemes are adaptive to the local statistics which in turn lead sensors to transmit a much better summary of local statistics to the fusion center with the same amount of resources compared to the conventional sampling. This makes event-triggered sampling schemes very convenient for decentralized change detection problems, see e.g., \cite{Li_15,Li14,Li16,Fellouris12}. Next, we explain the LCSH scheme.

\subsection{Level-Crossing Sampling with Hysteresis (LCSH)}

Implementing LCSH is quite simple. For each sensor, the amplitude axis is uniformly divided with spacing $\Delta$ and the corresponding levels are determined a priori. If the LR calculated at a sensor crosses a new upper/lower prespecified level, the sensor sends a sign bit 1/0 to the fusion center. If multiple levels are crossed at the same time, then, next to the sign bit, it sends 1/0 for each double/single additional crossings. As an example, let the most recently crossed level is $2 \Delta$ and the LR calculated for the new measurement is $6.4 \Delta$ at a sensor. Then, the sensor sends the bit sequence 110 to the fusion center to report four upward crossings. The LCSH scheme is illustrated in Figure \ref{fig:lcsh}.

\begin{figure}
\center
  \includegraphics[width=100mm]{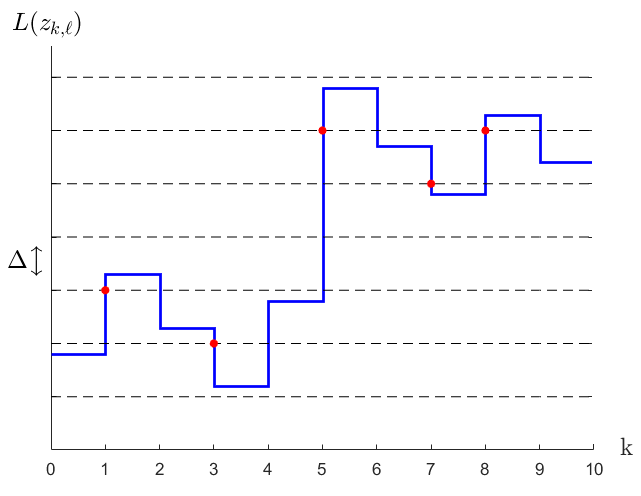}
  \vspace{-0.2cm}
\caption{LCSH scheme. The points indicating the pairs of sampling times and the corresponding crossed levels are shown (in red) in the figure.}
 \label{fig:lcsh}
\end{figure}

We assume that $\Delta$ is chosen the same for each sensor due to symmetry. At a sensor, say $\ell$, let the most recent sampling time be $t_{\ell,m-1}$ and the most recently crossed amplitude level in terms of $\Delta$ be $\eta_{\ell,m-1}$. Then, the next sampling time is determined as
\begin{gather} \nonumber
t_{\ell,m} = \min \big\{k \in \mathbb{N}: \, k > t_{\ell,m-1}, \, |L(z_{k,\ell})- \eta_{\ell,m-1} \Delta| \geq \Delta \big\},
\end{gather}
where the corresponding sign bit is calculated as
\begin{gather} \label{eq:sign_bit}
\zeta_{\ell,m} \triangleq \text{sign}\left(L(z_{t_{\ell,m},\ell}) - \eta_{\ell,m-1} \, \Delta\right),
\end{gather}
and the number of level crossings is determined as follows:
\begin{gather} \label{eq:level_cross}
\chi_{\ell,m} \triangleq \left \lfloor \frac{|L(z_{t_{\ell,m},\ell}) - \eta_{\ell,m-1} \, \Delta|}{\Delta} \right \rfloor \geq 1.
\end{gather}
Further, the number of bits transmitted from sensor $\ell$ to the fusion center is calculated as
\begin{gather} \nonumber
\varpi_{\ell,m} \triangleq \left\lceil \frac{\chi_{\ell,m}-1}{2}  \right\rceil + 1,
\end{gather}
and at both sensor $\ell$ and the fusion center, the most recently crossed level is updated as follows:
\begin{gather} \nonumber
\eta_{\ell,m} = \eta_{\ell,m-1} + \zeta_{\ell,m} \, \chi_{\ell,m}.
\end{gather}

The fusion center then approximates the LR of measurements obtained at sensor $\ell$ as $\eta_{\ell,m} \, \Delta$ until a new bit sequence is received from sensor $\ell$. Hence, the LR level reported to the fusion center is always in the $\pm \Delta$ range of the actual LR calculated at a sensor. That is, although the fusion center does not know the exact LR levels at sensors and thus has some approximation errors, such errors do not accumulate over time.

We note that in the limiting case, i.e., as $\Delta \rightarrow 0$, a decentralized setting with LCSH is equivalent to a centralized setting, however, this would require infinite number of bits for each transmission. Therefore, the level of $\Delta$ should be adjusted by considering the expected tradeoff between performance and resource usage. For a fair comparison between the conventional sampling and LCSH, we select $\Delta$ such that the average communication rates in both schemes are the same.

\begin{algorithm}[t]\small
\caption{\small The proposed procedure at sensor $\ell$}
\label{alg:sensor}
\baselineskip=0.5cm
\begin{algorithmic}[1]
\STATE Initialization: $k \gets 0$, $m \gets 1$, $\eta_{\ell,0} \gets 0$, ${\pmb{\varphi}}_k \gets {\pmb{\varphi}}^*, \forall k \geq 1$.
\WHILE {$k < \nu$}
    \STATE $k \gets k+1$.
    \IF {the centralized model is used}
        \STATE Send $z_{k,\ell}$ to the fusion center.
    \ELSIF{the decentralized model is used}
        \IF {the uniform-in-time sampling scheme is used}
            \IF {${\varphi}_{k,\ell,i} < L(z_{k,\ell}) \leq {\varphi}_{k,\ell,i+1}$}
                \STATE Send $u_{k,\ell} = i$ to the fusion center using $\lceil \log_2(U_\ell) \rceil$ bits.
            \ENDIF
        \ELSIF{the LCSH scheme is used}
            \IF {$|L(z_{k,\ell})- \eta_{\ell,m-1} \Delta| \geq \Delta$}
                \STATE Compute $\zeta_{\ell,m}$ and $\chi_{\ell,m}$ using \eqref{eq:sign_bit} and \eqref{eq:level_cross}, respectively.
                \STATE Send $\zeta_{\ell,m}$ and $\chi_{\ell,m}$ to the fusion center using $\varpi_{\ell,m}$ bits.
                \STATE $\eta_{\ell,m} \gets \eta_{\ell,m-1} + \zeta_{\ell,m} \, \chi_{\ell,m}$.
                \STATE $m \gets m+1$.
            \ENDIF
        \ENDIF
    \ENDIF
\ENDWHILE
\end{algorithmic}
\end{algorithm}

\subsection{Proposed Decentralized Tests using LCSH Scheme}

For both the multichart scheme and the uniform-prior scheme with LCSH, the procedures at a sensor are the same and summarized in Algorithm \ref{alg:sensor} (see lines 11-17). Next, we explain the procedures at the fusion center.

\subsubsection{Multichart Test with LCSH}

The multichart test with LCSH has a test structure as in \eqref{eq:tau_j} and \eqref{eq:multi_chart_v1}, and the recursion of sufficient statistics is as in \eqref{eq:recursion_pkn_reduced} where $\delta_{k,j}^{n}$ is calculated as follows:
\begin{gather} \label{eq:delta_reduced_LCS}
\delta_{k,j}^{n} = \prod_{i=1}^{n} \frac
{\mathbb{P}_1\left( L(z_{\pi_{i}}) = \eta_{\pi_{i}} \, \Delta  \right)}
{\mathbb{P}_0\left( L(z_{\pi_{i}}) = \eta_{\pi_{i}} \, \Delta  \right)},
\end{gather}
where $z_{\pi_{i}}$ denotes a generic observation at sensor $\pi_i$ and $\eta_{\pi_{i}}$ denotes the most recently crossed level in terms of $\Delta$ at sensor ${\pi_{i}}$ up to time $k$. Hence, at a time $k$, the fusion center approximates the LR of sensor $\ell$ as $\eta_{\ell} \, \Delta$. The procedure at the fusion center for the proposed multichart scheme with LCSH is summarized in Algorithm \ref{alg:multichart} (see lines 4-11 \& 17-25).

\subsubsection{Uniform-Prior Test with LCSH}

For the decentralized uniform-prior test with LCSH, the test structure is as in \eqref{eq:rare_change} and the recursion of sufficient statistics is as in \eqref{eq:recursion_pkn} where $\delta_{k}^{n}$ is calculated as follows:
\begin{gather} \nonumber
\delta_{k}^{n} = \bigg( \sum_{\Pi} \, \prod_{i=1}^{n} \mathbb{P}_1(L(z_{\pi_{i}}) = \eta_{\pi_{i}} \, \Delta)
 \prod_{i=n+1}^{L} \mathbb{P}_0(L(z_{\pi_{i}}) = \eta_{\pi_{i}} \, \Delta) \bigg) \\ \label{eq:delta_LCSH}
\times \left(\sum_{\Pi} \, \prod_{i=1}^{L} \mathbb{P}_0(L(z_{\pi_{i}}) = \eta_{\pi_{i}} \, \Delta) \right)^{-1}.
\end{gather}
Algorithm \ref{alg:uniprior} (see lines 9-14 \& 17-20) summarizes the procedure at the fusion center for the proposed uniform-prior test with LCSH.

\begin{algorithm}[t]\small
\caption{\small The proposed uniform-prior schemes: procedure at the fusion center}
\label{alg:uniprior}
\baselineskip=0.5cm
\begin{algorithmic}[1]
\STATE Initialization: $\beta \gets \log(1/(\rho \, \alpha))$, $k \gets 0$, $\eta_{\ell} \gets 0, \forall \ell \in \{1,2,\dots,L\}$, $p_{0}^{0} \gets {1}/{\rho}$, $p_{0}^{n} \gets 0, \forall n \in \{1, 2, \dots, L\}$, ${\pmb{\varphi}}_k \gets {\pmb{\varphi}}^*, \forall k \geq 1$.
\WHILE {$k < \nu$}
    \STATE $k \gets k+1$.
    \IF {the centralized model is used}
        \STATE Compute $\delta_k^n$ using \eqref{eq:delta_cent}, $\forall n \in \{1,2,\dots,L\}$.
    \ELSIF{the decentralized model is used}
        \IF {the uniform-in-time sampling scheme is used}
            \STATE Compute $\delta_k^n$ using \eqref{eq:delta}, $\forall n \in \{1,2,\dots,L\}$.
        \ELSIF{the LCSH scheme is used}
            \IF {a new bit sequence is received from any sensor $\ell \in \{1,2, \dots, L\}$}
                \STATE Determine the sign bit $\zeta_{\ell}$ and the number of level crossings $\chi_{\ell}$.
                \STATE $\eta_{\ell} \gets \eta_{\ell} + \zeta_{\ell} \, \chi_{\ell}$.
            \ENDIF
            \STATE Compute $\delta_k^n$ using \eqref{eq:delta_LCSH}, $\forall n \in \{1,2,\dots,L\}$.
        \ENDIF
    \ENDIF
    \STATE Compute $p_k^n$ using \eqref{eq:recursion_pkn}, $\forall n \in \{1,2,\dots,L\}$.
    \IF {$\log{\sum_{m=1}^{L} p_{k}^{m}} \geq \beta$}
        \STATE $\nu \gets k$.
    \ENDIF
\ENDWHILE
\STATE Declare the change and stop the procedure.
\end{algorithmic}
\end{algorithm}

\begin{algorithm}[t]\small
\caption{\small The proposed multichart schemes: procedure at the fusion center}
\label{alg:multichart}
\baselineskip=0.5cm
\begin{algorithmic}[1]
\STATE Initialization: $\beta \gets \log(1/(\rho \, \alpha))$, $k \gets 0$, $\eta_{\ell} \gets 0, \forall \ell \in \{1,2,\dots,L\}$, $p_{0,j}^{0} \gets {1}/{\rho}, \forall j \in \{1,2,\dots,L!\}$, $p_{0,j}^{n} \gets 0, \forall n \in \{1, 2, \dots, L\}, \forall j \in \{1,2,\dots,L!\}$, ${\pmb{\varphi}}_k \gets {\pmb{\varphi}}^*, \forall k \geq 1$.
\WHILE {$k < \nu$}
    \STATE $k \gets k+1$.
    \IF{the decentralized model and the LCSH scheme is used}
        \IF {a new bit sequence is received from any sensor $\ell \in \{1,2, \dots, L\}$}
            \STATE Determine the sign bit $\zeta_{\ell}$ and the number of level crossings $\chi_{\ell}$.
            \STATE $\eta_{\ell} \gets \eta_{\ell} + \zeta_{\ell} \, \chi_{\ell}$.
        \ENDIF
    \ENDIF
    \FORALL {$j = 1,2,\dots,L!$}
        \STATE Assume $\Pi = \Pi_j$.
        \IF {the centralized model is used}
            \STATE Compute $\delta_{k,j}^n$ using \eqref{eq:delta_reduced_cent}, $\forall n \in \{1,2,\dots,L\}$.
        \ELSIF{the decentralized model is used}
            \IF {the uniform-in-time sampling scheme is used}
                \STATE Compute $\delta_{k,j}^n$ using \eqref{eq:delta_reduced}, $\forall n \in \{1,2,\dots,L\}$.
            \ELSIF{the LCSH scheme is used}
                \STATE Compute $\delta_{k,j}^n$ using \eqref{eq:delta_reduced_LCS}, $\forall n \in \{1,2,\dots,L\}$.
            \ENDIF
        \ENDIF
        \STATE Compute $p_{k,j}^n$ using \eqref{eq:recursion_pkn_reduced}, $\forall n \in \{1,2,\dots,L\}$.
    \ENDFOR
    \IF {$\max_{\Pi_j} \, \log{\sum_{m=1}^{L} p_{k,j}^{m}} \geq \beta$}
        \STATE $\nu \gets k$.
    \ENDIF
\ENDWHILE
\STATE Declare the change and stop the procedure.
\end{algorithmic}
\end{algorithm}

\section{Change Detection Based on Online Estimation of the Change Propagation Pattern} \label{sec:online_estimate}

We have so far focused on detection-only schemes and observed that such schemes are computationally highly demanding and even infeasible to implement in real-time for large $L$ {(recall the $L!$ summations in the uniform-prior scheme (see \eqref{eq:delta}) and the $L!$ simultaneous tests in the multichart scheme (see \eqref{eq:multi_chart_v1}))}. This motivates us to propose a computationally effective scheme that jointly provides online estimates of the change propagation pattern and performs the change detection based on the estimated patterns.

\subsection{Online Estimation of the Change Pattern}

In our setup, the fusion center receives {either the exact measurements (centralized setting) or the quantized local LR statistics (decentralized setting)} from sensors through parallel communication channels. Hence, it is possible for the fusion center to compute the LR statistics for each sensor separately. In particular, the fusion center can compute and accumulate the log-likelihood ratios (LLRs) so as to obtain the individual (local) CUSUM statistics for each sensor. Using these statistics, the fusion center can then provide online estimates of the change propagation pattern, as detailed below.

Recall that in our problem, the change propagates through the network so that the sensors observe the change one by one at different (random) time-instants. For a sensor $\ell$ having observed the change until time $k$, we have $z_{k,\ell} \sim f_1$ and the expected value of the corresponding LLR at time $k$ is given by
\begin{gather} \label{eq:exp_llr_v0}
\mathbb{E}[\log(L(z_{k,\ell}))] = \int{f_1(z) \log\bigg(\frac{f_1(z)}{f_0(z)}\bigg) dz} = D(f_{1}, f_{0}) > 0.
\end{gather}
On the other hand, for a sensor $\ell'$ that has not observed the change yet, we have $z_{k,\ell'} \sim f_0$ and
\begin{gather}\label{eq:exp_llr}
\mathbb{E}[\log(L(z_{k,\ell'}))] = -D(f_{0}, f_{1}) < 0.
\end{gather}
Further, in a centralized setup, the individual CUSUM statistic for a sensor $\ell$ at time $k$, denoted with $C_{k,\ell}$, can be recursively updated as follows:
\begin{gather}\label{local_cusum}
C_{k,\ell} \gets \max\{0, C_{k-1,\ell} + \log(L(z_{k,\ell}))\},
\end{gather}
where $C_{0,\ell} = 0, \forall \ell \in \{1,\dots,L\}$. Based on \eqref{eq:exp_llr_v0}, \eqref{eq:exp_llr}, and \eqref{local_cusum}, if the sensor $\ell$ has observed the change before the sensor $\ell'$, we have $\mathbb{E}[C_{k,\ell}] > \mathbb{E}[C_{k,\ell'}]$, i.e., the CUSUM statistics are expected to be larger for sensors having observed the change earlier. On the other hand, if both $\ell$ and $\ell'$ have not observed the change until time $k$, then the comparison between $C_{k,\ell}$ and $C_{k,\ell'}$ is not informative.

For the sensors having not observed the change yet, since the expected value of their LLRs is negative (see \eqref{eq:exp_llr}), their individual CUSUM statistics are not expected to reach high values (see \eqref{local_cusum}). Hence, we propose that the fusion center employs a thresholding procedure for the individual CUSUM statistics at each time $k$ and separates the sensors into two mutually exclusive and disjoint groups: the first group $\mathcal{S}_{1,k}$ consists of the sensors with the CUSUM statistics exceeding a threshold $\xi > 0$ (if any) and the second group $\mathcal{S}_{2,k}$ consists of the sensors with the CUSUM statistics less than $\xi$ (if any). Clearly, the fusion center estimates that the change propagates through $\mathcal{S}_{1,k}$ before $\mathcal{S}_{2,k}$. Next, we explain the estimation of the change pattern within each group.

Since the sensors, having not observed the change yet, have the same pre-change measurement pdf $f_0$, they are all identical from the change detection perspective. Hence, we estimate the change propagation pattern between sensors in the set $\mathcal{S}_{2,k}$ uniformly randomly. On the other hand, for the sensors belonging to $\mathcal{S}_{1,k}$ that have reliably high CUSUM statistics (based on the chosen threshold $\xi$), we estimate the change propagation pattern based on the ordering between their individual CUSUM statistics. In other words, denoting the estimated change pattern at time $k$ by $\hat{\Pi}_k = [\hat{\pi}_1, \hat{\pi}_2, \dots, \hat{\pi}_L]$, $\hat{\pi}_1$ is estimated as the sensor having the largest CUSUM statistic, $\hat{\pi}_2$ is estimated as the sensor having the second largest CUSUM statistic, and so on (assuming $\hat{\pi}_1, \hat{\pi}_2 \in \mathcal{S}_{1,k}$). {Alg.~\ref{alg:estimate} summarizes the online estimation of the change pattern in both the centralized and decentralized settings. Note that in the decentralized settings, we propose to compute the individual CUSUM statistics by accumulating the logarithm of the quantized LRs computed based on the received messages from sensors (see the lines 1-20).}

\subsection{Change Detection Based on Estimated Change Patterns}

We now propose change detection schemes based on the online estimation of the change propagation pattern. In this method, a change pattern is estimated at each time among the $L!$ possibilities and hence the computational complexity of the corresponding detection scheme is significantly reduced. Recall that in the multichart test, $L!$ feasible tests are simultaneously employed where the test for a particular pattern $\Pi_j$ is given by \eqref{eq:tau_j} and the recursion of the test statistics is given by \eqref{eq:recursion_pkn_reduced}. Here, we only replace $\Pi_j$ with the estimated pattern $\hat{\Pi}_k$ at each time $k$ and obtain the test based on the online estimated patterns as follows:
\begin{gather} \label{eq:new_tau_j}
\hat{\nu}(\beta) = \inf \bigg\{k \in \mathbb{N}: \log{\sum_{m=1}^{L} \hat{p}_{k}^{m}} \geq \beta \bigg\},
\end{gather}
where
\begin{gather} \label{eq:new_recursion_pkn_reduced}
\hat{p}_{k}^{n} = \hat{\delta}_{k}^{n} \, \frac{1-\rho_{\pi_{n},\pi_{n+1}}}{1-\rho} \, \bigg(\sum_{m=0}^{n} \hat{p}_{k-1}^{m} \, e_{m}^{n}\bigg).
\end{gather}
As before, we present the centralized and the decentralized versions of this test and for the different versions, only $\hat{\delta}_{k}^{n}$ changes. In particular, $\hat{\delta}_{k}^{n}$ is computed for the centralized version as
\begin{gather} \label{eq:new_delta_reduced_cent}
\hat{\delta}_{k}^{n} =  \prod_{i=1}^{n} \frac{f_1(z_{k,\hat{\pi}_{i}})}{f_0(z_{k,\hat{\pi}_{i}})},
\end{gather}
for the decentralized version with US as
\begin{gather} \label{eq:new_delta_reduced}
\hat{\delta}_{k}^{n} =  \prod_{i=1}^{n} \frac{\mathbb{P}_1\big({\varphi}_{\hat{\pi}_{i}, u_{k,\hat{\pi}_{i}}} < L(z_{\hat{\pi}_{i}}) \leq {\varphi}_{\hat{\pi}_{i},(u_{k,\hat{\pi}_{i}}+1)}\big)}{\mathbb{P}_0\big({\varphi}_{\hat{\pi}_{i}, u_{k,\hat{\pi}_{i}}} < L(z_{\hat{\pi}_{i}}) \leq {\varphi}_{\hat{\pi}_{i},(u_{k,\hat{\pi}_{i}}+1)}\big)},
\end{gather}
and for the decentralized version with LCSH as follows:
\begin{gather} \label{eq:new_delta_reduced_LCS}
\hat{\delta}_{k}^{n} = \prod_{i=1}^{n} \frac
{\mathbb{P}_1\left( L(z_{\hat{\pi}_{i}}) = \eta_{\hat{\pi}_{i}} \, \Delta  \right)}
{\mathbb{P}_0\left( L(z_{\hat{\pi}_{i}}) = \eta_{\hat{\pi}_{i}} \, \Delta  \right)},
\end{gather}
{where $z_{\hat{\pi}_{i}} \sim f_m$, $m \in \{0,1\}$ denotes a generic observation at sensor $\hat{\pi}_i$ and $L(z_{\hat{\pi}_{i}})$ denotes the corresponding LR. We summarize the proposed change pattern estimation-based detection procedures at the fusion center in Alg.~\ref{alg:estimate}. Moreover, as before, we summarize the proposed procedures at the sensor nodes in Alg.~\ref{alg:sensor}.}

Finally, we note that the threshold $\xi$ affects the change detection performance. In particular, choosing a small $\xi$ may increase the false alarm rate of the proposed estimation-based detector. This is because during the pre-change period where the measurement pdfs are $f_0$ across all sensors, a small (close to zero) $\xi$ does not uniformly randomize the sensors and it particularly favors the (random) measurements that seem to be distributed with $f_1$. On the other hand, choosing a large $\xi$ may cause the change pattern estimation algorithm to react slowly in distinguishing sensors with reliably high CUSUM statistics and hence the detection delays can be increased. In practice, a good $\xi$ can be determined based on an offline simulation.

\begin{algorithm}[t]\small

\caption{\small {Online estimation of the change pattern $\Pi$ and the estimation-based detection procedures at the fusion center}}
\label{alg:estimate}
\baselineskip=0.5cm
\begin{algorithmic}[1]
\STATE Initialization: $k \gets 0$, $\eta_{\ell} \gets 0, \forall \ell \in \{1,\dots,L\}$, $\hat{p}_{0}^{0} \gets {1}/{\rho}$, $\hat{p}_{0}^{n} \gets 0, \forall n \in \{1, 2, \dots, L\}$, $\pmb{\varphi}_{\ell} \gets {\pmb{\varphi}}^*$, $\forall \ell \in \{1,\dots,L\}$, $C_{0,\ell} \gets 0, \forall \ell \in \{1,\dots,L\}$.
\WHILE {$\log{\sum_{n=1}^{L} \hat{p}_{k}^{n}} < \beta$}
    \STATE $k \gets k+1$
    \IF {the centralized model is used}
        \STATE $C_{k,\ell} \gets  \max\{0, C_{k-1,\ell} + \log(L(z_{k,\ell}))\}$, $\forall \ell \in \{1,\dots,L\}$
    \ELSIF{the decentralized model is used}
        \IF {the US scheme is used}
            \STATE $C_{k,\ell} \gets  \max\left\{0, C_{k-1,\ell} + \log\left(\frac{\mathbb{P}_1\big({\varphi}_{\ell, u_{k,\ell}} < L(z_{\ell}) \leq {\varphi}_{\ell,(u_{k,\ell}+1)} \big)}{\mathbb{P}_0\big({\varphi}_{\ell, u_{k,\ell}} < L(z_{\ell}) \leq {\varphi}_{\ell,(u_{k,\ell}+1)} \big)}\right)\right\}$, $\forall \ell \in \{1,\dots,L\}$
        \ELSIF{the LCSH scheme is used}
            \IF {a new bit sequence is received from any sensor $\ell \in \{1,\dots, L\}$}
                \STATE Determine the sign bit $\zeta_{\ell}$ and the number of level crossings $\chi_{\ell}$.
                \STATE $\eta_{\ell} \gets \eta_{\ell} + \zeta_{\ell} \, \chi_{\ell}$.
            \ENDIF
            \STATE $C_{k,\ell} \gets  \max\{0, C_{k-1,\ell} + \log(\eta_{\ell} \, \Delta)\}$, $\forall \ell \in \{1,\dots,L\}$
        \ENDIF
    \ENDIF
    \STATE Determine $\mathcal{S}_{1,k} \triangleq \{\ell: C_{k,\ell} \geq \xi\}$ and $\mathcal{S}_{2,k} \triangleq \{\ell: C_{k,\ell} < \xi\}$.
    \STATE Sort the indices $\ell \in \mathcal{S}_{1,k}$ in decreasing order according to $\{C_{k,\ell}: \ell \in \mathcal{S}_{1,k}\}$ and denote the resulting vector by $\hat{\Pi}_{1,k}$.
    \STATE Permute the indices $\ell \in \mathcal{S}_{2,k}$ uniformly randomly and denote the resulting vector by $\hat{\Pi}_{2,k}$.
    \STATE $\hat{\Pi}_k \gets [\hat{\Pi}_{1,k}, \, \hat{\Pi}_{2,k}]$
    \IF {the centralized model is used}
        \STATE Compute $\hat{\delta}_{k}^{n}$ using \eqref{eq:new_delta_reduced_cent}, $\forall n \in \{1,\dots,L\}$.
    \ELSIF{the decentralized model is used}
        \IF {the US scheme is used}
            \STATE Compute $\hat{\delta}_{k}^{n}$ using \eqref{eq:new_delta_reduced}, $\forall n \in \{1,\dots,L\}$.
        \ELSIF{the LCSH scheme is used}
            \STATE Compute $\hat{\delta}_{k}^{n}$ using \eqref{eq:new_delta_reduced_LCS}, $\forall n \in \{1,\dots,L\}$.
        \ENDIF
    \ENDIF
    \STATE Update $\hat{p}_{k}^{n}$ using \eqref{eq:new_recursion_pkn_reduced}, $\forall n \in \{1,\dots,L\}$.
\ENDWHILE
\STATE $\hat{\nu}(\beta) \gets k$, declare the change and stop the procedure.
\end{algorithmic}
\end{algorithm}

\section{Numerical Results} \label{sec:sim}

In this section, we evaluate performance of the proposed tests summarized in Table \ref{table:prop_tests} and compare them with some other tests in the literature via numerical examples. Firstly, the proposed decentralized tests with US are compared with (i) the decentralized version of the order-1 asymptotically optimal test designed for the known change propagation pattern \cite{Raghavan10}, (ii) the decentralized order-1 asymptotically optimal test under the assumption of simultaneous change point for all sensors (denoted with mismatched test) \cite{Tartakovsky05}, and (iii) the test in (i) using measurements of only one sensor which observes the change first (denoted with single sensor test). Then, the proposed centralized tests are compared with the centralized versions of (i), (ii) and (iii). Finally, the proposed decentralized tests are compared with the proposed centralized tests and performance improvement obtained via LCSH is illustrated. Throughout this section, we consider a system consisting of three sensors, i.e., $L=3$. The geometric parameter for the first change point is taken as $\rho = 0.01$, and the tests are evaluated with different inter-sensor change propagation parameters. Particularly, $\lambda$ takes values of $0.01$, $0.1$, $0.3$, and $0.9$. Moreover, for the test based on the online change pattern estimation, we choose $\xi = 3$.

\begin{table*}[t]
    \centering
    \begin{tabular}{ | p{7cm} | l | l | l |}
    \hline
    Description & Stopping Rule & Recursion of Suff. Stat. & $\delta_{k}^{n}/\delta_{k,j}^{n}/\hat{\delta}_{k}^{n}$ \\ \hline \hline
    Centralized uniform-prior test & \eqref{eq:rare_change} & \eqref{eq:recursion_pkn} & \eqref{eq:delta_cent}  \\ \hline
    Decentralized uniform-prior test w/ US & \eqref{eq:rare_change} & \eqref{eq:recursion_pkn} & \eqref{eq:delta} \\ \hline
    Decentralized uniform-prior test w/ LCSH & \eqref{eq:rare_change} & \eqref{eq:recursion_pkn} & \eqref{eq:delta_LCSH} \\ \hline
    Centralized multichart test & \eqref{eq:multi_chart_cent} & \eqref{eq:recursion_pkn_reduced} & \eqref{eq:delta_reduced_cent} \\ \hline
    Decentralized multichart test w/ US & \eqref{eq:multi_chart_v1} & \eqref{eq:recursion_pkn_reduced} & \eqref{eq:delta_reduced} \\ \hline
    Decentralized multichart test w/ LCSH & \eqref{eq:multi_chart_v1} & \eqref{eq:recursion_pkn_reduced} & \eqref{eq:delta_reduced_LCS} \\
    \hline
    Centralized estimation-based test  & \eqref{eq:new_tau_j} & \eqref{eq:new_recursion_pkn_reduced} & \eqref{eq:new_delta_reduced_cent} \\ \hline
    Decentralized estimation-based test w/ US & \eqref{eq:new_tau_j} & \eqref{eq:new_recursion_pkn_reduced} & \eqref{eq:new_delta_reduced} \\ \hline
    Decentralized estimation-based test w/ LCSH & \eqref{eq:new_tau_j} & \eqref{eq:new_recursion_pkn_reduced} & \eqref{eq:new_delta_reduced_LCS} \\
    \hline
    \end{tabular}
    \vspace{-0.05cm}
    \caption{Summary of the proposed tests.}
    \label{table:prop_tests}
    \vspace{-0.45cm}
\end{table*}

Considering the stringent energy constraints of sensor nodes and for the sake of simplicity, we use US with binary quantization, i.e., each sensor sends one bit at a time to report its LR. The transmitted bit from sensor $\ell$ to the fusion center at time $k$ is determined as follows:
\begin{equation} \nonumber
    u_{k,\ell} =
    \begin{cases}
      1, & \text{if} ~~ L(z_{k,\ell}) > \varphi, \\
      0, & \text{else},
    \end{cases}
\end{equation}
where $\varphi$ denotes the single finite quantization threshold. As an example, let the densities be $f_0 \sim \mathcal{N}(0,1)$ and $f_1 \sim \mathcal{N}(\mu,1)$. Then, $D(f_1, f_0) = {\mu^2}/{2}$ and $L(z_{k,\ell}) = \exp{(z_{k,\ell} \, \mu - {\mu^2}/{2})}$. Let ${\varrho \triangleq {\log{(\varphi)}}/{\mu} + {\mu}/{2}}$ be the threshold on the level of the observations. The transmitted bit is then determined as follows:
\begin{equation} \nonumber
    u_{k,\ell} =
    \begin{cases}
      1, & \text{if} ~~ z_{k,\ell} > \varrho, \\
      0, & \text{else}.
    \end{cases}
\end{equation}

The pmfs on sensor messages can then be calculated using \eqref{eq:induced_pmf0} and \eqref{eq:induced_pmf1} and the K-L information distance between the pmfs is obtained as follows:
\begin{align} \nonumber
D(P_{1,\ell}^{u}, P_{0,\ell}^{u}) &= (1 - Q(\varrho - \mu)) \log\left(\frac{1 - Q(\varrho - \mu)}{1 - Q(\varrho)}\right) \\ \nonumber
 &+ \, Q(\varrho - \mu) \log\left(\frac{Q(\varrho - \mu)}{Q(\varrho)} \right),
\end{align}
where $Q(y) = \frac{1}{\sqrt{2 \pi}} \int_{y}^{\infty} {\exp{(-\frac{{x^2}}{2})} \, dx}$.
Let $\mu = 1$ as an example. The threshold maximizing $D(P_{1,\ell}^{u}, P_{0,\ell}^{u})$ is then calculated as ${\varrho^* = 0.7942}$ and the corresponding K-L information distance is $D(P_{1}^{u^*}, P_{0}^{u^*}) = 0.3186$.

Fig. \ref{fig:decentralized} shows the tradeoff curves between the ADD and PFA of the decentralized tests with 1-bit US. The slope of the first-order asymptote is ${1}/({L \, D(P_{1}^{u^*}, P_{0}^{u^*}) + |\log(1 - \rho)|}) = 1.0354$ (cf. \eqref{eq:decent_perf}). According to Corollary 1, the decentralized multichart test with US is order-1 asymptotically optimal if the condition given in \eqref{eq:cond_opt_dc} holds. Note that \eqref{eq:cond_opt_dc} is satisfied if $\lambda \geq 0.8074$. Hence, the condition holds for $\lambda = 0.9$ and does not hold for $\lambda = 0.3$, $\lambda = 0.1$, and $\lambda = 0.01$. We, however, observe through Fig. \ref{fig:decentralized} that especially for $\lambda = 0.3$, the first-order asymptotic optimality holds, even for the uniform-prior test. This implies that
the sufficient condition \eqref{eq:cond_opt_dc} for the first-order asymptotic optimality might be quite strong.

\begin{figure*}[t]
\center
  \includegraphics[width=160mm]{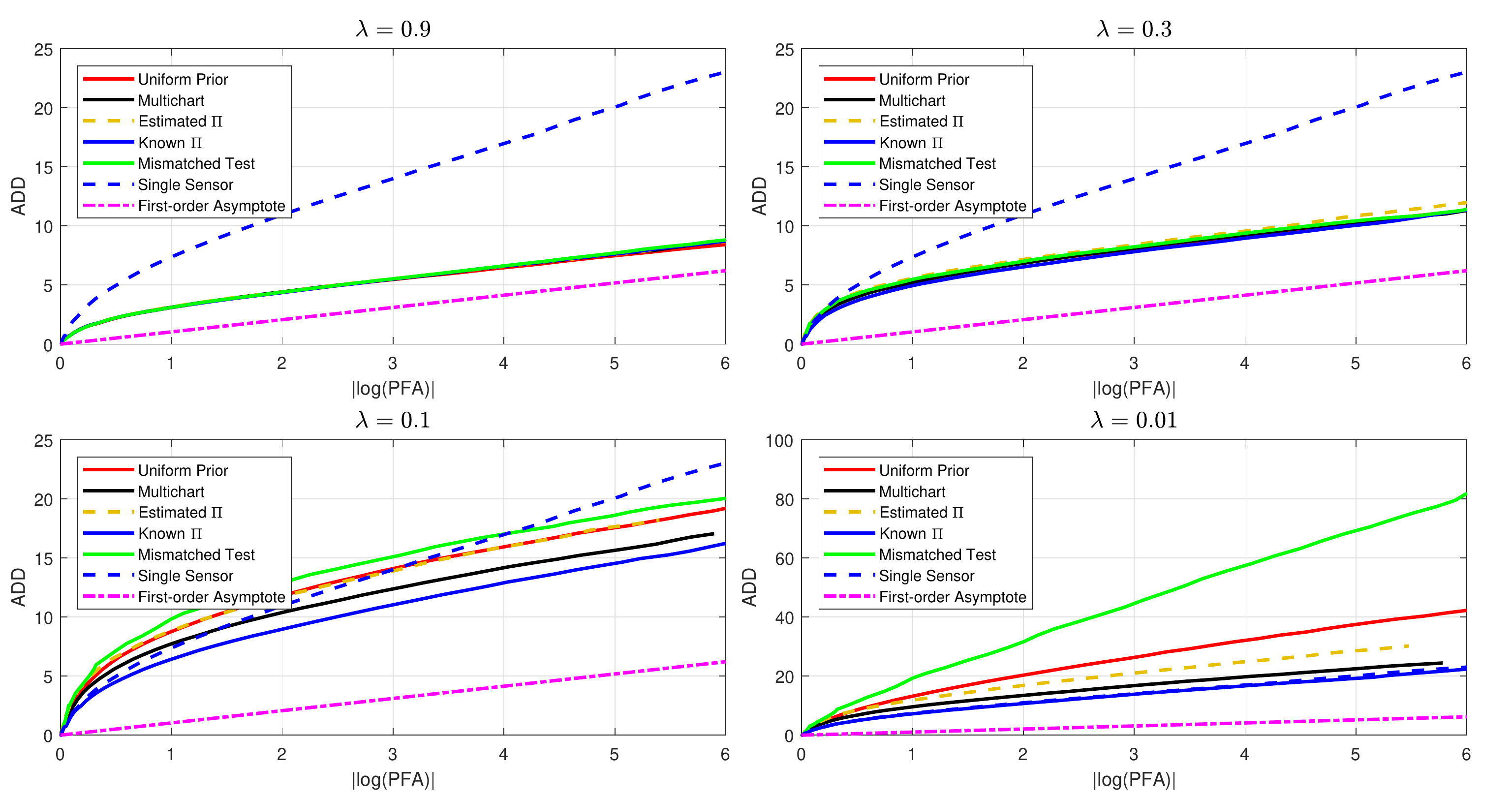}
  \vspace{-0.2cm}
\caption{Comparison of the proposed decentralized tests with US with some existing tests in the literature.}
 \label{fig:decentralized}
 \vspace{-0.4cm}
\end{figure*}

Fig. \ref{fig:centralized} shows the tradeoff curves for the centralized tests. The sufficient condition for the asymptotic optimality given in \eqref{eq:cond_opt} holds for $\lambda \geq 0.6706$ so the condition is satisfied for $\lambda = 0.9$ and not satisfied for $\lambda = 0.3$, $\lambda = 0.1$, and $\lambda = 0.01$. Note that $D(f_1, f_0) = 0.5$ is greater than $D(P_{1}^{u^*}, P_{0}^{u^*})$, as expected. The slope of the first-order asymptote is equal to $0.6622$ that is smaller than the slope of the first-order asymptote in the decentralized setting, which is $1.0354$. Therefore, for a given PFA level, ADD of the centralized tests are significantly smaller than ADD of the decentralized tests with 1-bit US.

\begin{figure*}[t]
\center
  \includegraphics[width=160mm]{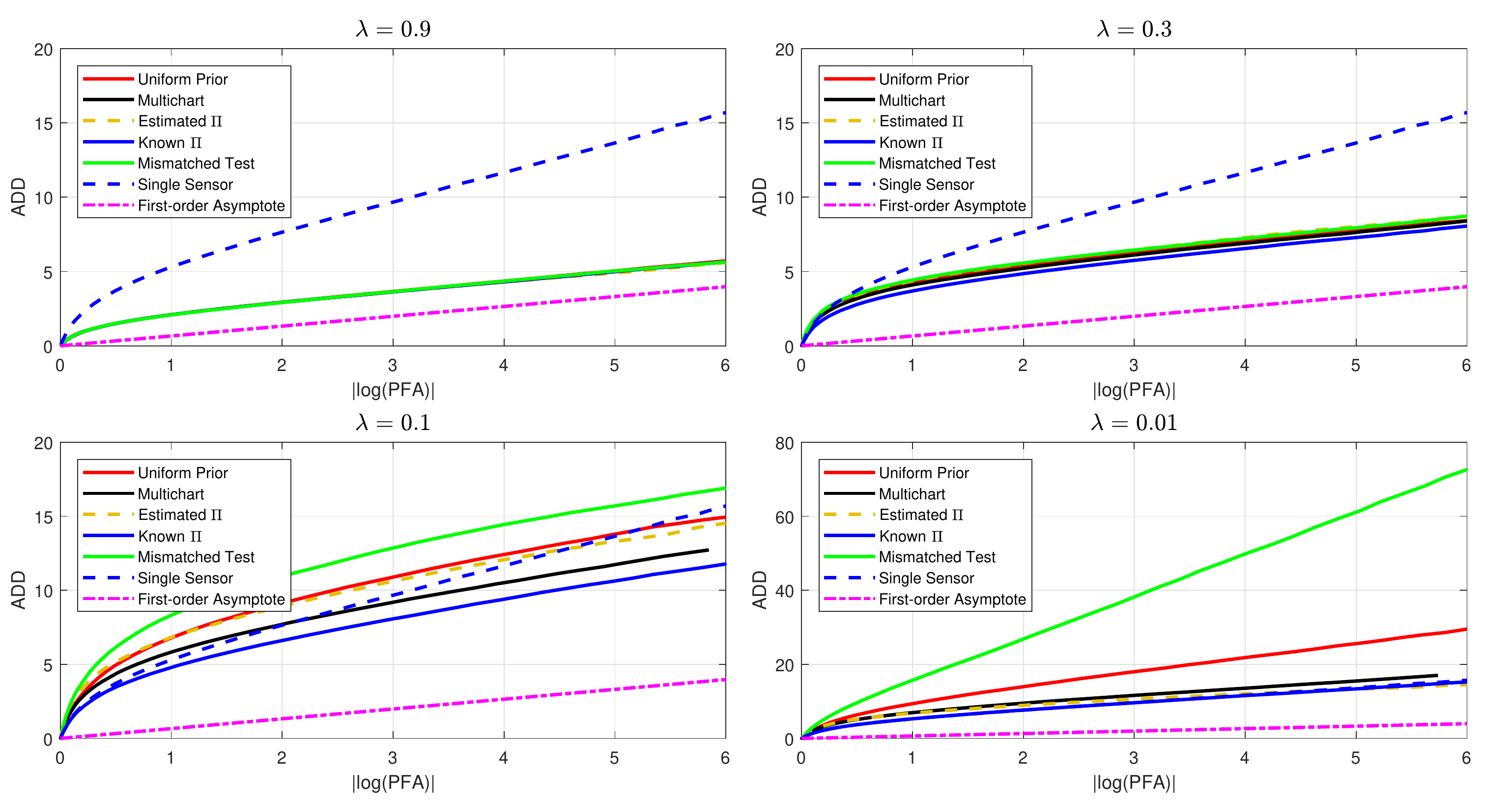}
  \vspace{-0.2cm}
\caption{Comparison of the proposed centralized tests with some existing tests in the literature.}
 \label{fig:centralized}
 \vspace{-0.6cm}
\end{figure*}

For both the centralized and decentralized settings, we observe through Fig. \ref{fig:decentralized} and Fig. \ref{fig:centralized} that the tests with the known change pattern always have the smallest ADD for the same levels of PFA, hence have the best performance. In fact, the known-$\Pi$ case can be considered as a performance upper bound for the proposed tests. Then through the figures, we observe that the multichart tests in general perform better than the uniform-prior tests and the tests based on online change pattern estimation. Moreover, the proposed tests get closer to the performance upper bound as $\lambda$ increases. Also, we observe that although the computational complexity is significantly reduced for the test based on online pattern estimation, its performance is comparable to the multichart and uniform-prior tests. Hence, especially for large networks consisting of many sensors (large $L$), this computationally effective detection scheme can be reliably used.

Further, we observe that the performance of the mismatched tests improves as $\lambda$ increases and the performance of the single sensor tests improves as $\lambda$ decreases. This is expected since in the mismatched scheme, $\lambda = 1$ is assumed and the single sensor scheme corresponds to the case where $\lambda = 0$. Our proposed tests considerably outperform the mismatched tests as $\lambda$ gets closer to zero and the single sensor tests as $\lambda$ gets closer to one. Moreover, the average detection delays of the proposed tests decrease as $\lambda$ increases. This is due to the fact that positive effect of the observational diversity on the performance is more clearly seen for bigger $\lambda$, i.e., as $\lambda$ gets bigger, more sensors contribute to the change detection.

We now compare the proposed centralized and decentralized tests. Note that a decentralized test can at most achieve performance of its centralized counterpart. Fig. \ref{fig:multichart_lcsh} show the tradeoff curves for the multichart tests in the centralized setting and in the decentralized setting with US and LCSH schemes. We observe through the figure that due to 1-bit coarse quantization, there is a significant performance loss in the decentralized tests with US compared to the centralized tests. Moreover, for all $\lambda$ levels, we observe that the decentralized tests with LCSH show considerable performance improvement compared to those with conventional sampling. Note that $\Delta$ is computed via offline simulations such that the average communication rates in both US and LCSH schemes are nearly equal to each other for the same levels of PFA. Hence, LCSH can significantly improve the performance of the decentralized tests using the same amount of communication resources. We further note that for the uniform-prior test and the test based on the online estimated change patterns, we observe similar results, i.e., LCSH schemes closely perform to their centralized counterparts and significantly outperform the conventional US schemes. Hence, we only present the results for the multichart test in Fig. \ref{fig:multichart_lcsh} as a representative for all proposed tests.


\begin{figure*}[t]
\center
  \includegraphics[width=160mm]{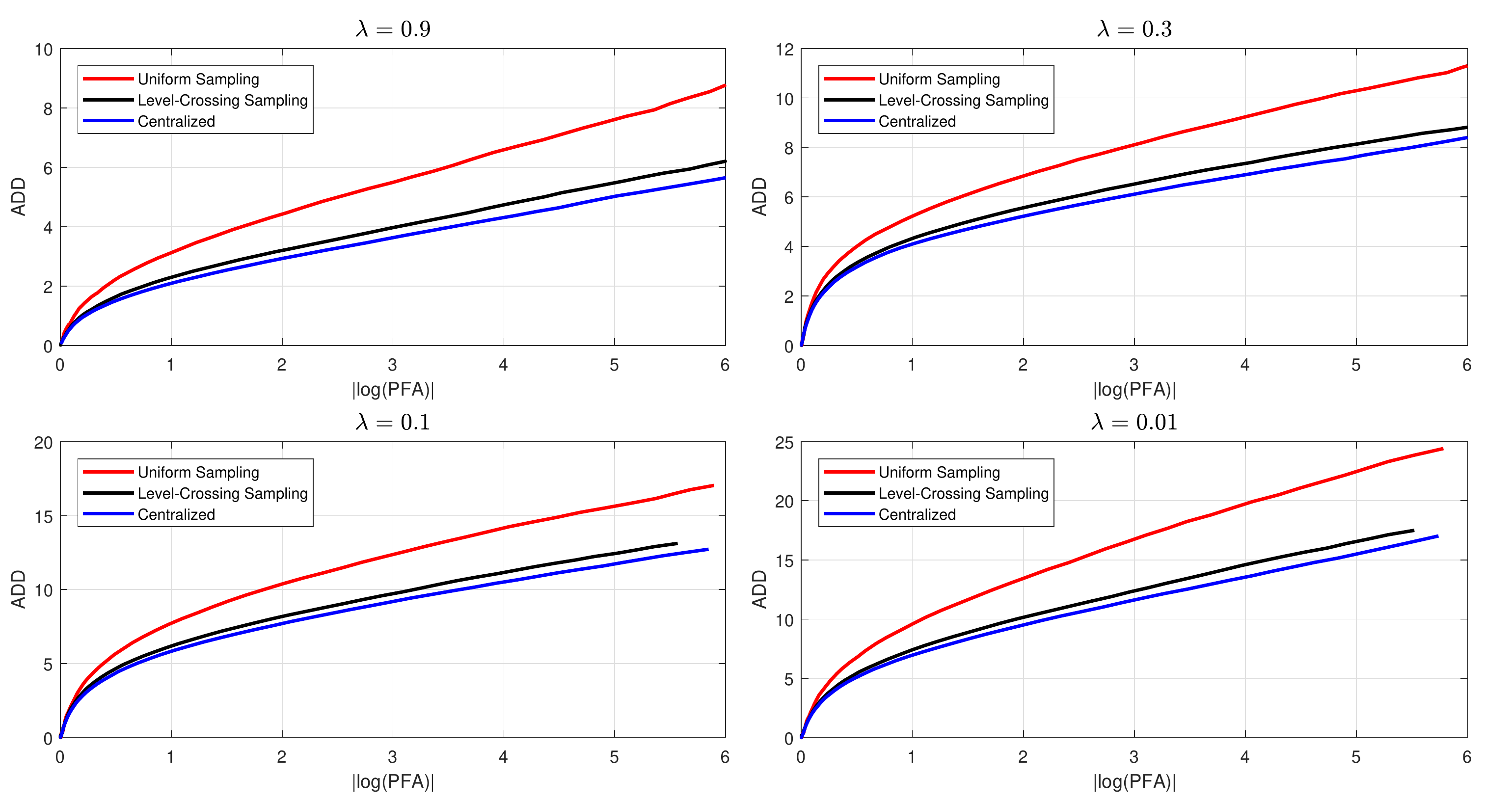}
  \vspace{-0.2cm}
\caption{Comparison of the decentralized multichart tests with the centralized multichart test.}
 \label{fig:multichart_lcsh}
 \vspace{-0.6cm}
\end{figure*}

\section{Concluding Remarks} \label{sec:conc}

We have studied a Bayesian sequential change detection problem with unknown change propagation pattern in both centralized and decentralized settings. We have considered the cases where the change propagation pattern is a random variable with a known prior distribution, and where it is an unknown non-random parameter, and studied both of the corresponding schemes. Using the DP framework, we have provided insights into the optimal stopping rules. In the practical rare change regime, we have shown that the optimal stopping rule is a threshold test and its false alarm probability can be upper bounded with an easy-to-compute threshold. Moreover, for the case where the change pattern is considered a non-random quantity, we have provided sufficient conditions for the first-order asymptotic optimality of the corresponding test. Also, with the purpose of obtaining computationally feasible test for large networks consisting of many sensors, we have presented a detection scheme that is employed based on the online estimates of the change propagation pattern. Further, in the decentralized setting, to reduce the performance loss due to quantization, we have proposed to use an event-triggered sampling scheme called LCSH and shown its advantages compared to the conventional US scheme. We have provided numerical examples to illustrate the near-optimal performance of the proposed tests.

\section*{Acknowledgment}

The authors would like to thank Dr. Vasanthan Raghavan for his useful comments.

\appendix

\subsection{Proof of Proposition 1} \label{sec:proof_thm1}

\begin{proof}
The proof is similar to \cite[Proof of Theorem 2]{Veeravalli01} with some modifications. In particular, we extend the proof by considering the general setup where change-points of sensors may not be the same. Considering the finite-horizon DP equations given in \eqref{eq:finiteDPk_v1}, let ${R_k \triangleq \mathbb{E}\, [\tilde{J}_{k}^{T}(I_{k}) \,|\, I_{k-1}]}$ where the expectation is taken with respect to the measurements made at the sensor nodes at time $k$, i.e., ${z_{k,1}, \dots, z_{k,L}}$. The aim is to minimize $R_k$ by choosing $\pmb{\phi}_k = \{\phi_{k,1}, \dots, \phi_{k,L}\}$ optimally. We fix all quantization functions at time $k$ except one of them, say $\phi_{k,i}$, and minimize $R_k$ with respect to only $\phi_{k,i}$. Hence, we derive the structure of all person-by-person optimal (p.b.p.o.) solutions. Since the globally optimal solution is also p.b.p.o., it also follows this structure.

We can rewrite $R_k$ as follows:
\begin{align} \label{eq:iter_exp}
R_k &= \mathbb{E}\, \big[\mathbb{E}\, [\tilde{J}_{k}^{T}(u_{k,1}, \dots, u_{k,L}, I_{k-1}) \,|\, \pmb{\Gamma}, I_{k-1}] \,|\, I_{k-1}\big],
\end{align}
where $\pmb{\Gamma}=[\Gamma_1, \dots, \Gamma_L]$ denotes the collection of the change-points of sensors. In \eqref{eq:iter_exp}, the inner expectation is with respect to $z_{k,1}, \dots, z_{k,i-1}, z_{k,i+1}, \dots, z_{k,L} \, | \, \pmb{\Gamma}, I_{k-1}$ and the outer expectation is with respect to $z_{k,i}, \pmb{\Gamma} | I_{k-1}$. The conditioning on $z_{k,i}$ in the inner expectation is dropped due to independence of the observations given $\pmb{\Gamma}$. Result of the inner expectation is a function of $u_{k,i}$, $\pmb{\Gamma}$ and $I_{k-1}$. Let this function be denoted with $Y(u_{k,i}, \pmb{\Gamma}, I_{k-1})$, which does not depend on $u_{k,\ell}$'s except $u_{k,i}$ since we have assumed that all $\phi_{k,\ell}$'s except $\phi_{k,i}$ are fixed and under this assumption, when we take expectation with respect to all $z_{k,\ell}$'s except $z_{k,i}$, all $u_{k,\ell}$'s except $u_{k,i}$ disappear. Then, we have
\begin{align} \nonumber
R_k &= \mathbb{E}\, [Y(u_{k,i}, \pmb{\Gamma}, I_{k-1}) \,|\, I_{k-1}] \\ \label{eq:iter_exp2}
&= \mathbb{E}\, \big[\mathbb{E}\, [Y(u_{k,i}, \pmb{\Gamma}, I_{k-1}) \,|\, z_{k,i}, I_{k-1}] \,|\, I_{k-1}\big].
\end{align}

In \eqref{eq:iter_exp2}, the inner expectation is with respect to $\pmb{\Gamma} | z_{k,i}, I_{k-1}$ and the outer expectation is with respect to $z_{k,i} | I_{k-1}$. By definition, $\Gamma_i, i \in \{1, \dots, L\}$ takes integer values in the range $[1,\infty)$. Let the range of $\pmb{\Gamma}$ be denoted with $\Sigma$ and partitioned into two parts at time $k$ as $\Sigma = \Sigma_{k,i}^{1} \cup \Sigma_{k,i}^{0}$ where $\Sigma_{k,i}^{1} \triangleq \{ \pmb{\Gamma}: \Gamma_i \leq k \}$ and $\Sigma_{k,i}^{0} \triangleq \{ \pmb{\Gamma}: \Gamma_i > k \}$.

Then,
\begin{gather} \nonumber
R_k = \mathbb{E}\, \left[\sum_{\pmb{\gamma} \in \Sigma_{k,i}^{1}} Y(u_{k,i}, \pmb{\gamma}, I_{k-1}) \mathbb{P}(\pmb{\Gamma} = \pmb{\gamma} \,|\, z_{k,i}, I_{k-1}) + \sum_{\pmb{\gamma} \in \Sigma_{k,i}^{0}} Y(u_{k,i}, \pmb{\gamma}, I_{k-1}) \mathbb{P}(\pmb{\Gamma} = \pmb{\gamma} \,|\, z_{k,i}, I_{k-1}) \, \right] \\  \nonumber
= \mathbb{E}\, \left[\sum_{\pmb{\gamma} \in \Sigma_{k,i}^{1}} Y(u_{k,i}, \pmb{\gamma}, I_{k-1}) \frac{f_1(z_{k,i}) \mathbb{P}(\pmb{\Gamma} = \pmb{\gamma} \,|\, I_{k-1})}{\mathbb{P}(z_i = z_{k,i} \,|\, I_{k-1})} + \sum_{\pmb{\gamma} \in \Sigma_{k,i}^{0}} Y(u_{k,i}, \pmb{\gamma}, I_{k-1}) \frac{f_0(z_{k,i}) \mathbb{P}(\pmb{\Gamma} = \pmb{\gamma} \,|\, I_{k-1})}{\mathbb{P}(z_i = z_{k,i} \,|\, I_{k-1})} \, \right] \\ \label{eq:iter_exp6}
= \mathbb{E}\, \bigg[ \frac{1}{\mathbb{P}(z_i = z_{k,i} \,|\, I_{k-1})} \bigg(f_1(z_{k,i}) W^{1}(u_{k,i}, I_{k-1}) +  f_0(z_{k,i}) W^{0}(u_{k,i}, I_{k-1})\bigg) \bigg],
\end{gather}
where
\begin{gather} \nonumber
W^{1}(u_{k,i}, I_{k-1}) \triangleq \sum_{\pmb{\gamma} \in \Sigma_{k,i}^{1}} Y(u_{k,i}, \pmb{\gamma}, I_{k-1}) \mathbb{P}(\pmb{\Gamma} = \pmb{\gamma} \,|\, I_{k-1}),
\end{gather}
\begin{gather} \nonumber
W^{0}(u_{k,i}, I_{k-1}) \triangleq \sum_{\pmb{\gamma} \in \Sigma_{k,i}^{0}} Y(u_{k,i}, \pmb{\gamma}, I_{k-1}) \mathbb{P}(\pmb{\Gamma} = \pmb{\gamma} \,|\, I_{k-1}),
\end{gather}
and $z_i$ denotes a generic measurement at sensor $i$.

Then, the goal is to minimize the expectation \eqref{eq:iter_exp6} with respect to $u_{k,i} = \phi_{k,i}(z_{k,i})$, which is equivalent to minimizing inside the expectation in \eqref{eq:iter_exp6} almost everywhere. That is,
\begin{align} \nonumber
\phi_{k,i}(z_{k,i}) &= \arg \min_{u_{k,i} \, \in \, \{0, \dots, U_i-1\}} \frac{1}{\mathbb{P}(z_i = z_{k,i} \,|\, I_{k-1})} \bigg(
f_1(z_{k,i}) W^{1}(u_{k,i}, I_{k-1}) + f_0(z_{k,i}) W^{0}(u_{k,i}, I_{k-1}) \bigg) \\ \label{eq:opt_quant_fin}
&= \arg \min_{u_{k,i} \, \in \, \{0, \dots, U_i-1\}} \bigg( \frac{f_1(z_{k,i})}{f_0(z_{k,i})} W^{1}(u_{k,i}, I_{k-1})
+  W^{0}(u_{k,i}, I_{k-1}) \bigg).
\end{align}
As a special case, if $\frac{f_1(z_{k,i})}{f_0(z_{k,i})} = \infty$, then
\begin{gather} \label{eq:opt_quant_inf}
\phi_{k,i}(z_{k,i}) = \arg \min_{u_{k,i} \, \in \, \{0, \dots, U_i-1\}} W^{1}(u_{k,i}, I_{k-1}).
\end{gather}
Note that ${f_1(z_{k,i})}/{f_0(z_{k,i})}$ is the likelihood ratio of $z_{k,i}$. Based on \eqref{eq:opt_quant_fin} and \eqref{eq:opt_quant_inf}, we observe that $\phi_{k,i}(z_{k,i})$ can be expressed as a minimum of linear functions of the likelihood ratio, which defines an interval of likelihood ratio to be mapped to each ${u_{k,i} \, \in \, \{0, \dots, U_i-1\}}$. This implies that the structure of optimal local message functions in the finite-horizon is in the form of likelihood ratio quantizers. We also observe that $\phi_{k,i}(z_{k,i})$ is a function of $I_{k-1}$ and therefore the thresholds of the quantizer depend on $I_{k-1}$. This completes the proof.
\end{proof}

\subsection{Proof of Lemma 2} \label{sec:proof_lemma2}

\begin{proof}

(a) Since $J_{T}^{T}(\mathbf{q}_T) = q_{T}^{0}$ is an affine function of $\mathbf{q}_T$, it is clearly concave in $\mathbf{q}_T$. Moreover, we observe that for any choice of $\pmb{\phi}_{T}$,
\begin{align} \nonumber
{\mathbb{E}\,[J_{T}^{T}(\mathbf{q}_T)\,|\,I_{T-1}]}
&= {\mathbb{E}\,[q_{T}^{0}\,|\,I_{T-1}]} \\ \nonumber
&= {\mathbb{E}\,[\mathbb{P}(S_{T} = 0 \, | \, \mathbf{u}_T ,I_{T-1})\,|\,I_{T-1}]} \\ \nonumber
&= {\mathbb{E}\,\bigg[\frac{\mathbb{P}(\mathbf{u}_T \,|\, S_{T} = 0, I_{T-1})}{\mathbb{P}(\mathbf{u}_T \,|\, I_{T-1})} \mathbb{P}(S_{T} = 0 \,|\, I_{T-1}) \,\big{|}\,I_{T-1}\bigg]} \\ \nonumber
&= {\mathbb{E}\,\bigg[\frac{\mathbb{P}(\mathbf{u}_T \,|\, S_{T} = 0, I_{T-1})}{\mathbb{P}(\mathbf{u}_T \,|\, I_{T-1})} \mathbb{P}(S_{T-1} = 0, \Gamma_{\pi_1} \neq T \,|\, I_{T-1}) \,\big{|}\,I_{T-1}\bigg]} \\ \nonumber
&= {\mathbb{E}\,\bigg[\frac{\mathbb{P}(\mathbf{u}_T \,|\, S_{T} = 0, I_{T-1})}{\mathbb{P}(\mathbf{u}_T \,|\, I_{T-1})} \underbrace{\mathbb{P}(S_{T-1} = 0 \,|\, I_{T-1})}_{q_{T-1}^{0}} \underbrace{\mathbb{P}(\Gamma_{\pi_1} \neq T \,|\, S_{T-1} = 0, I_{T-1})}_{(1-\rho)} \,\big{|}\,I_{T-1}\bigg]} \\ \nonumber
&= (1-\rho) q_{T-1}^{0} \sum_{\mathbf{u}_T} \frac{\mathbb{P}(\mathbf{u}_T \,|\, S_{T} = 0, I_{T-1})}{\mathbb{P}(\mathbf{u}_T \,|\, I_{T-1})} \mathbb{P}(\mathbf{u}_T \,|\, I_{T-1}) \\ \nonumber
&= (1-\rho) q_{T-1}^{0} \underbrace{\sum_{\mathbf{u}_T} \mathbb{P}(\mathbf{u}_T \,|\, S_{T} = 0, I_{T-1})}_{1} \\ \nonumber
&= (1-\rho) q_{T-1}^{0}.
\end{align}

Then, using \eqref{A_k_T_v2}, we have
\begin{align} \nonumber
A_{T-1}^{T}(\mathbf{q}_{T-1})
&= \min_{\pmb{\phi}_{T}} {\mathbb{E}\,[J_{T}^{T}(\mathbf{q}_T)\,|\,I_{T-1}]} \\ \nonumber
&= (1 - \rho) q_{T-1}^{0},
\end{align}
which is a concave function of $\mathbf{q}_{T-1}$. Since
\begin{align} \nonumber
J_{T-1}^{T}(\mathbf{q}_{T-1})
&= \min\big\{ q_{T-1}^{0}, c \, (1 - q_{T-1}^{0}) + (1 - \rho) q_{T-1}^{0} \big\}, 
\end{align}
i.e., the minimum of two affine functions of $\mathbf{q}_{T-1}$, $J_{T-1}^{T}(\mathbf{q}_{T-1})$ is a concave function of $\mathbf{q}_{T-1}$.

We now use an induction argument to show that $J_{k}^{T}(\mathbf{q}_{k})$ and $A_{k}^{T}(\mathbf{q}_{k})$ are concave functions of $\mathbf{q}_{k}$ for all $k \in \{0, 1, \dots, T-2\}$. Suppose that for any $k \in \{0, 1, \dots, T-2\}$, $J_{k}^{T}(\mathbf{q}_{k})$ is a concave function of $\mathbf{q}_{k}$. This implies that $J_{k}^{T}(\mathbf{q}_{k})$ can be expressed as infimum of a set $\mathcal{M}$ of affine functions of $\mathbf{q}_{k}$ as follows \cite{Boyd04}:
\begin{gather} \nonumber
J_{k}^{T}(\mathbf{q}_{k}) = \inf \big\{f_m(\mathbf{q}_{k}) \in \mathcal{M}: f_m(\mathbf{q}_{k}) = \pmb{\kappa}_m^{\mathsf{T}} \, \mathbf{q}_{k} + \iota_m, \, f_m(\mathbf{q}_{k}) \geq J_{k}^{T}(\mathbf{q}_{k}) \text{ for all } \mathbf{q}_{k} \in \mathcal{Q} \big\},
\end{gather}
where $\pmb{\kappa}_m = [\kappa_m^0, \kappa_m^1, \dots, \kappa_m^L]$ and $\iota_m$ are some constants defining an affine function $f_m \in \mathcal{M}$ of $\mathbf{q}_{k}$, and ${\mathsf{T}}$ represents the vector transpose. Then,
\begin{align} \nonumber
A_{k-1}^{T}(\mathbf{q}_{k-1})
&= \min_{\pmb{\phi}_{k}} {\mathbb{E}\,[J_{k}^{T}(\mathbf{q}_k)\,|\,I_{k-1}]} \\ \nonumber
&= \min_{\pmb{\phi}_{k}} \sum_{\mathbf{u}_k} \bigg(\inf_{f_m \in \mathcal{M}} \pmb{\kappa}_m^{\mathsf{T}} \, \frac{\mathbf{g}_{k}(\mathbf{u}_k, \pmb{\phi}_{k}, \mathbf{q}_{k-1})}{h_k(\mathbf{u}_k, \pmb{\phi}_{k}, \mathbf{q}_{k-1})} + \iota_m \bigg) \, h_k(\mathbf{u}_k, \pmb{\phi}_{k}, \mathbf{q}_{k-1}) \\ \label{eq:A_k_proof}
&= \min_{\pmb{\phi}_{k}} \inf_{f_m \in \mathcal{M}} \sum_{\mathbf{u}_k} \pmb{\kappa}_m^{\mathsf{T}} \, {\mathbf{g}_{k}(\mathbf{u}_k, \pmb{\phi}_{k}, \mathbf{q}_{k-1})}
+ \iota_m \, {h_k(\mathbf{u}_k, \pmb{\phi}_{k}, \mathbf{q}_{k-1})}.
\end{align}
It is clear from \eqref{eq:recursion_g_v2} and \eqref{eq:p_uk_ik_1} that $g_k^n, \forall n \in \{0,1,\dots,L\}$ and $h_k$ are affine functions of $\mathbf{q}_{k-1}$. In \eqref{eq:A_k_proof}, we observe that $A_{k-1}^{T}(\mathbf{q}_{k-1})$ is expressed as minimum and infimum over a set of affine functions of $\mathbf{q}_{k-1}$, hence it is a concave function of $\mathbf{q}_{k-1}$. Furthermore, through \eqref{eq:finiteDPk_v2}, since the sum of concave functions is concave and the minimum of concave functions is concave, $J_{k-1}^{T}(\mathbf{q}_{k-1})$ is a concave function of $\mathbf{q}_{k-1}$. This completes the proof of part (a).

(b) Since $J_{T}^{T}(\mathbf{q}_T) = q_{T}^{0}$ is nonnegative, through the structure of $J_{k}^{T}(\mathbf{q}_{k})$ given in \eqref{eq:finiteDPk_v2}, using simple induction arguments, it is easy to see that $J_{k}^{T}(\mathbf{q}_{k})$ and $A_{k}^{T}(\mathbf{q}_{k})$ are nonnegative for all $0 \leq k \leq T$. Moreover, due to \eqref{eq:finiteDPk_v2}, $J_{k}^{T}(\mathbf{q}_{k}) \leq q_{k}^{0} \leq 1$ for all $0 \leq k \leq T$. Further, if $q_{k}^{0} = 0$ for any $k$, then $0 \leq J_{k}^{T}(\mathbf{q}_{k}) \leq q_{k}^{0}$ implies that $J_{k}^{T}(\mathbf{q}_{k}) = 0$. Moreover, since $0 \leq J_{k+1}^{T}(\mathbf{q}_{k+1}) \leq 1$, we have $0 \leq \mathbb{E}\,[J_{k+1}^{T}(\mathbf{q}_{k+1})\,|\,I_{k}] \leq 1$. Then, since $A_{k}^{T}(\mathbf{q}_{k}) = \min_{\pmb{\phi}_{k+1}} \mathbb{E}\,[J_{k+1}^{T}(\mathbf{q}_{k+1})\,|\,I_{k}]$, we have $0 \leq A_{k}^{T}(\mathbf{q}_{k}) \leq 1$. Further, if $q_{k}^{0} = 0$, using the recursion of sufficient statistics (cf. \eqref{eq:recursion_g_v2}-\eqref{eq:recursion_v2}), it is straightforward to check that $q_{k+1}^{0} = 0$, which implies that $J_{k+1}^{T}(\mathbf{q}_{k+1}) = 0$. Then, since $0 \leq A_{k}^{T}(\mathbf{q}_{k}) = \min_{\pmb{\phi}_{k+1}} \mathbb{E}\,[J_{k+1}^{T}(\mathbf{q}_{k+1})\,|\,I_{k}] = 0$, we have $A_{k}^{T}(\mathbf{q}_{k}) = 0$ if $q_{k}^{0} = 0$.
\end{proof}

\subsection{Proof of Proposition 3} \label{sec:proof_prop1}

\begin{proof}

(a) The proof is similar to \cite[Theorem 3]{Veeravalli93} with some modifications. For completeness, we provide the proof. Firstly, we notice that (i) $J \in \mathcal{S}$, (ii) $A(\mathbf{q}) = \min_{\pmb{\phi} \, \subset \, \pmb{\Phi}} K_J(\pmb{\phi}, \mathbf{q})$, and (iii) $J(\mathbf{q})$ is a fixed point of the mapping, i.e., $\omega J(\mathbf{q}) = J(\mathbf{q})$. Assume that there exists another fixed point $B \in \mathcal{S}$ which is different than $J$, i.e., $\omega B(\mathbf{q}) = B(\mathbf{q})$ and $B \neq J$. Let $\pmb{\phi}^* = \arg \min_{\pmb{\phi} \, \subset \, \pmb{\Phi}} K_B(\pmb{\phi}, \mathbf{q})$. Then, we define a stopping rule $\mathcal{T} \triangleq \min\{k \in \mathbb{N}: q_k^{0} \leq c \, (1 - q_k^{0}) +  K_B(\pmb{\phi}^*, \mathbf{q}_k)\}$. We choose an initial value for the sufficient statistic $\mathbf{q}_0 \in \mathcal{Q}$ and define $\{\mathbf{q}_k, k \geq 1\}$ using the recursion of the sufficient statistics given in \eqref{eq:recursion_v7}. Since $B$ is a fixed point of the mapping $\omega$, then for all $\mathbf{q}_k, k \in \{0,1,\dots,\mathcal{T}-1\}$, we have
\begin{align} \nonumber
B(\mathbf{q}_k) &= \omega B(\mathbf{q}_k) \\ \label{eq:asd}
&= \min \{q_k^{0}, \, c \, (1 - q_k^{0}) + K_B(\pmb{\phi}^*, \mathbf{q}_k) \} \\ \label{eq:qwe}
&= \min \{q_k^{0}, \, c \, (1 - q_k^{0}) + \mathbb{E}\,[B(\mathbf{q}_{k+1}) \, | \, I_k] \} \\ \nonumber
&= c \, (1 - q_k^{0}) +  \mathbb{E}\,[B(\mathbf{q}_{k+1}) \, | \, I_k],
\end{align}
and for $k = \mathcal{T}$, we have $B(\mathbf{q}_{\mathcal{T}}) = \omega B(\mathbf{q}_{\mathcal{T}}) = q_{\mathcal{T}}^{0}$. Note that \eqref{eq:asd} is obtained by the definition of $\omega$ given in \eqref{eq:mapp} and \eqref{eq:qwe} is obtained from \eqref{eq:asd} using \eqref{eq:K_B}. Then, by the definition of the stopping rule $\mathcal{T}$, we have the following equations:
\begin{align}\nonumber
  B(\mathbf{q}_0) &= c \, (1 - q_0^{0}) +  \mathbb{E}\,[B(\mathbf{q}_1)], \\ \nonumber
  B(\mathbf{q}_1) &= c \, (1 - q_1^{0}) +  \mathbb{E}\,[B(\mathbf{q}_2) \,|\, I_1], \\ \nonumber
  &\vdots \\ \nonumber
  B(\mathbf{q}_{\mathcal{T}-1}) &= c \, (1 - q_{\mathcal{T}-1}^{0}) +  \mathbb{E}\,[B(\mathbf{q}_\mathcal{T}) \,|\, I_{\mathcal{T}-1}], \\ \nonumber
  B(\mathbf{q}_{\mathcal{T}}) &= q_{\mathcal{T}}^{0}.
\end{align}

We now substitute these equations backwards to obtain the value of $B(\mathbf{q}_0)$. Note that for any $k \geq 1$, we have
\begin{align} \nonumber
\mathbb{E}\,[q_k^0 \,|\, I_{k-1}] &= \mathbb{E}\,[\mathbb{P}(S_k = 0 \,|\, \mathbf{u}_k, I_{k-1}) \,|\, I_{k-1}] \\ \nonumber
&= \sum_{\mathbf{u}_k} \frac{\mathbb{P}(\mathbf{u}_k \,|\, S_k = 0, I_{k-1}) \mathbb{P}(S_k = 0 \,|\, I_{k-1})}{\mathbb{P}(\mathbf{u}_k \,|\, I_{k-1})} \mathbb{P}(\mathbf{u}_k \,|\, I_{k-1}) \\ \label{eq:trm}
&= \mathbb{P}(S_k = 0 \,|\, I_{k-1}).
\end{align}
We firstly substitute $B(\mathbf{q}_{\mathcal{T}})$ into $B(\mathbf{q}_{\mathcal{T}-1})$ and using \eqref{eq:trm}, we have
\begin{align}\nonumber
B(\mathbf{q}_{\mathcal{T}-1}) &= c \, (1 - q_{\mathcal{T}-1}^{0}) + \mathbb{P}(S_{\mathcal{T}} = 0 \,|\, I_{\mathcal{T}-1}).
\end{align}
Then, we substitute $B(\mathbf{q}_{\mathcal{T}-1})$ into $B(\mathbf{q}_{\mathcal{T}-2})$ and so on. Recalling that $I_0$ is an empty set, we finally obtain
\begin{align}\nonumber
  B(\mathbf{q}_0) &= \mathbb{P}(S_{\mathcal{T}} = 0) + c \, \sum_{k=0}^{\mathcal{T}-1} \mathbb{P}(S_k \geq 1),
\end{align}
which is equal to the Bayes risk given in \eqref{eq:risk_v2} if the stopping rule is chosen as $\mathcal{T}$ and the local message functions are chosen as $\pmb{\phi}^*$. According to the DP arguments, the minimum Bayes risk is equal to $J(\mathbf{q}_0)$, which implies $B(\mathbf{q}_0) \geq J(\mathbf{q}_0), \forall \mathbf{q}_0 \in \mathcal{Q}$.

If we can show that the reverse inequality is also true, then the proof will be completed. To this end, we firstly notice that for all $\mathbf{q} \in \mathcal{Q}$, $B(\mathbf{q}) = \omega B(\mathbf{q}) \leq q^{0} = J_T^T(\mathbf{q})$ for any $T \geq 1$. We now use an induction argument. Suppose that for any $m \leq T-1$, we have $J_m^T(\mathbf{q}) \geq B(\mathbf{q})$. Then, using \eqref{eq:finiteDPk_v2}, \eqref{A_k_T_v2}, and \eqref{eq:K_B}, we obtain
\begin{align}\nonumber
  J_{m-1}^T(\mathbf{q}) &= \min\{q^{0}, c \, (1 - q^{0}) + \min_{\pmb{\phi} \, \subset \, \pmb{\Phi}} K_{J_m^T}(\pmb{\phi}, \mathbf{q}) \} \\
   &\geq \min\{q^{0}, c \, (1 - q^{0}) + \min_{\pmb{\phi} \, \subset \, \pmb{\Phi}} K_{B}(\pmb{\phi}, \mathbf{q}) \} \\
   &= \omega B(\mathbf{q}) = B(\mathbf{q}).
\end{align}

Hence, for all $m \leq T-1$ and for all $\mathbf{q} \in \mathcal{Q}$, we have $J_{m}^T(\mathbf{q}) \geq B(\mathbf{q})$. Taking the limit, we have $\lim_{T \rightarrow \infty} J_{m}^T(\mathbf{q}) = J(\mathbf{q}) \geq B(\mathbf{q})$. Therefore, $J(\mathbf{q}) = B(\mathbf{q}), \forall \mathbf{q} \in \mathbf{Q}$. This contradicts with the initial assumption that $J \neq B$. Hence, there exists only one fixed point of the mapping $\omega$. Since $J(\mathbf{q})$ is a fixed point, it is the unique fixed point.

(b) Let $\pmb{\phi}^* = \arg \min_{\pmb{\phi} \, \subset \, \pmb{\Phi}} K_J(\pmb{\phi}, \mathbf{q})$. If we solve the problem \eqref{eq:opt_prob} by restricting the local message functions to $\pmb{\phi}^*$ for all $k$, following the same solution methodology, we end up with the following Bellman equation:
\begin{align} \nonumber
\bar{J}(\mathbf{q}) &= \min\{q^{0}, \, c \, (1 - q^{0}) + K_{\bar{J}}(\pmb{\phi}^*, \mathbf{q})\}, 
\end{align}
where $\bar{J}(\mathbf{q})$ denotes the corresponding infinite-horizon cost-to-go function. Clearly, $\bar{J}(\mathbf{q})$ is a fixed point of the mapping $\omega$, which is defined in part (a). Since in part (a), we have proved that there is a unique fixed point of the mapping $\omega$, $\bar{J}(\mathbf{q}) = J(\mathbf{q})$ must be true. Hence, there is no loss of optimality if the local message functions are chosen to be fixed over time.
\end{proof}

\subsection{Proof of Proposition 4-(a)} \label{sec:proof_rare}

\begin{proof}

The proof is almost same with \cite[Proof of Theorem 2]{Raghavan10}. Hence, we only provide a sketch. In the rare change regime, as previously stated, the optimal sensor thresholds are constant. Then, we assume that the sensor thresholds are initially chosen at time $k=0$.
Then, using \eqref{eq:temp}, we express the finite-horizon DP equations given in \eqref{eq:finiteDPk_v2} in terms of $\mathbf{p}_k$ as follows:
\begin{gather} \label{eq:finiteDPT_v3} \nonumber
J_{T}^{T}(\mathbf{p}_T) = \frac{1}{1 + \rho \, \sum_{m=1}^{L} p_{T}^{m}},
\end{gather}
and for $0 \leq k < T$,
\begin{gather} \label{eq:finiteDPk_v3} \nonumber
J_{k}^{T}(\mathbf{p}_{k}) = \min\bigg\{ \frac{1}{1 + \rho \, \sum_{m=1}^{L} p_{k}^{m}},
\frac{c\,\rho\,\sum_{m=1}^{L} p_{k}^{m}}{1 + \rho \, \sum_{m=1}^{L} p_{k}^{m}} + A_{k}^{T} (\mathbf{p}_{k}) \bigg\},
\end{gather}
where
\begin{align} \nonumber
A_{k}^{T} (\mathbf{p}_{k})
&= \mathbb{E}\,[J_{k+1}^{T}(\mathbf{p}_{k+1})\,|\,I_{k}].
\end{align}
We then define the following functions:
\begin{gather} \label{eq:Phi_k} \nonumber
\Phi_k \triangleq \frac{1}{1 + \rho \, \sum_{m=1}^{L} p_{k}^{m}} - J_{k}^{T}(\mathbf{p}_{k}), \\ \label{eq:Psi_k} \nonumber
\Psi_k \triangleq A_{k}^{T}(\mathbf{p}_{k}) - \frac{1-\rho}{1 + \rho \, \sum_{m=1}^{L} p_{k}^{m}}.
\end{gather}
Based on these definitions, we have
\begin{gather} \nonumber
\Psi_k = - \mathbb{E}\,[\Phi_{k+1}\,|\,I_{k}].
\end{gather}
Then, for $0 \leq k < T$,
\begin{gather} \nonumber
J_{k}^{T}(\mathbf{p}_{k}) = \min\bigg\{ \frac{1}{1 + \rho \, \sum_{m=1}^{L} p_{k}^{m}}, \,
\frac{1 - \rho + c\,\rho\,\sum_{m=1}^{L} p_{k}^{m}}{1 + \rho \, \sum_{m=1}^{L} p_{k}^{m}} + \Psi_k \bigg\}.
\end{gather}
The test structure then takes the following form:
\begin{gather} \label{eq:test_rare}
\tau = \min \bigg\{k: \sum_{m=1}^{L} p_{k}^{m} \geq \frac{1}{c} \, \frac{1 - \frac{\Psi_k}{\rho}}{1 + \frac{\Psi_k}{c}} \bigg\}.
\end{gather}
Following the identical arguments given in \cite[Proof of Theorem 2]{Raghavan10}, it can be shown that (i) $\Phi_k$ and $\Psi_k$ are upper bounded by a function of $\rho$ for all $k \leq T$ and this function converges to zero as $\rho \rightarrow 0$, (ii) ${\Psi_k}/{\rho} \rightarrow 0$ as $\rho \rightarrow 0$ for all $k \leq T$, where the statements (i) and (ii) are both independent of the value of $T$. Then, taking the limit of \eqref{eq:test_rare} as $\rho \rightarrow 0$ and taking the logarithm of both sides of the inequality, we obtain the test structure as given in \eqref{eq:rare_change}.
\end{proof}

\bibliographystyle{IEEEtran}
\bibliography{refs,sequential_detection_bib}

\end{document}